\documentclass[a4paper]{article}

\usepackage[english]{babel}
  \usepackage[utf8]{inputenc} 
\usepackage{amsmath, amssymb, amsthm}
\usepackage{graphicx}
\usepackage{floatrow}
\usepackage{chngcntr}
\usepackage{authblk}
\usepackage{fullpage}
\usepackage{tikz}
\usetikzlibrary{decorations.pathmorphing}
\usetikzlibrary{patterns}
\usepackage{todonotes}
\usepackage{mathrsfs}
\usepackage{comment}
\usepackage{verbatim}
\usepackage{sidecap}
\usepackage{soul}

\usepackage{caption}
\usepackage{subcaption}

\usepackage{xpatch}
\makeatletter
\xpatchcmd{\@thm}{\thm@headpunct{.}}{\thm@headpunct{}}{}{}
\makeatother

\newcommand{\R}{\mathbb{R}}
\newcommand{\Prb}{\phi}
\newcommand{\pr}{\mathbb{P}}

\newcommand{\N}{\mathbb{N}}

\newcommand{\Z}{\mathbb{Z}}
\newcommand{\cc}{\leftrightarrow}

\newcommand{\cco}[1]{\overset{#1}{\cc}}

\newcommand{\abs}[1]{ \left \vert  #1  \right \vert}

\newcommand{\Rt}{ \hat{\mathcal{R}}^*_\sim}

\newcommand{\nn}{\mathbf{n}}
\newcommand{\mm}{\mathbf{m}}

\theoremstyle{plain}
\newtheorem{theorem}{Theorem}[section]

\newtheorem{lemma}[theorem]{Lemma}
\newtheorem{claim}[theorem]{Claim}

\newtheorem{corollary}[theorem]{Corollary}

\newtheorem{proposition}[theorem]{Proposition}

\title{Mass scaling of the near-critical 2D Ising model using random currents}
\author{Frederik Ravn Klausen\footnote{QMATH, University of Copenhagen, klausen@math.ku.dk} , Aran Raoufi\footnote{aranraoufi@gmail.com} }
\date{\today}

\begin{document}
\maketitle

\begin{abstract}
\noindent 
We examine the Ising model at its critical temperature with an external magnetic field $h a^{\frac{15}{8}}$ on $a\Z^2$ for $a,h >0$. 
A new proof of exponential decay of the truncated two-point correlation functions is presented. It is proven that the mass (inverse correlation length) is of the order of $h^\frac{8}{15}$ in the limit $h \to 0$. This was previously proven with CLE-methods in \cite{cjn20}. Our new proof uses instead the random current representation of the Ising model and its backbone exploration. The method further relies on recent couplings to the random cluster model  \cite{ADTW19}  as well as a near-critical RSW-result for the random cluster model  \cite{DCM20}.  
\end{abstract} 
\section{Introduction} 
The square lattice Ising model \cite{Isi25} suggested by Lenz \cite{Lenz20}   is the archetypal statistical physics model undergoing an order/disorder phase transition. It has been subject of  intense study in the past century \cite{Gri06, DC17}, starting with Periels' proof of the existence of a phase transition \cite{Per36} and Onsager's calculation of the free energy \cite{Ons44}. 
The rigorous understanding of the critical two-dimensional Ising model has advanced tremendously in the past decade starting with the breakthroughs \cite{Smi10, CheSmi12} and with the subsequent works (see, for example, \cite{DumSmi12a}).

One of the questions that remained unsolved until recently is obtaining the speed of the decay of the truncated correlations in the near-critical two-dimensional Ising model. 
For $a \in (0,1]$ and $h > 0$ the near critical regime is defined to be the Ising measure on the lattice $a \mathbb{Z}^2$ with the parameter $\beta = \beta_c (\mathbb Z^2)$ and external field $a^{15/8}\, h$. We denote the corresponding correlation functions with $\langle . \rangle_{a,h}$. The following theorem is proved in \cite{cjn20} using the scaling limit of the FK-Ising model which was proved to exist in \cite{camia2016planar} and its connections to the conformal loop ensemble \cite{CCK17}. See also the review \cite{CJNrev}.

\begin{theorem} \label{main}
There exists $B_0, C_0 \in (0, \infty)$ such that for any $a \in (0,1 \rbrack$ and $h > 0$ with $h a^{\frac{15}{8}} \leq 1$,
\begin{align*}
0 \leq \langle \sigma_x  \sigma_y \rangle_{a,h} - \langle \sigma_x  \rangle_{a,h} \langle \sigma_y  \rangle_{a,h} \leq C_0 a^{\frac{1}{4}} \abs{x-y}^{-\frac{1}{4}} e^{- B_0 h^{\frac{8}{15}}\abs{x-y}}. 
\end{align*}
\noindent Accordingly, for $a=1$ the result on $\Z^2$ is that for any $h \in \lbrack 0, 1) $,
\begin{align*}
\langle \sigma_y  \sigma_x \rangle_{1, h} -\langle \sigma_y  \rangle_{1, h} \langle  \sigma_x \rangle_{1, h}  \leq C_0  \abs{x -y}^{-\frac{1}{4}} e^{- B_0 h^{\frac{8}{15}} \abs{x-y}}. 
\end{align*}
\end{theorem}

\noindent  In this paper we prove Theorem \ref{prop3} from which we can deduce Theorem \ref{main}.

\begin{theorem}\label{prop3}
For any $h > 0$ and $a \leq 1$ there are functions $C(h)> 0$ and $m(h) > 0$ independent of $a>0$ such that for any $x,y \in a\Z^2$ it holds that
\begin{align*}
 \langle \sigma_x  \sigma_y \rangle_{a,h} - \langle \sigma_x  \rangle_{a,h} \langle \sigma_y  \rangle_{a,h} \leq C(h) a^{\frac{1}{4}} \abs{x-y}^{-\frac{1}{4}} e^{-m(h) \abs{x -y}}. 
\end{align*}
\end{theorem}

\noindent The proof uses first a partial exploration of the backbone of random currents,  then a recent coupling between the random current measure with sources and the random cluster model \cite{ADTW19}. The proof utilises a new result that extends a result on crossing probabilities for the critical random cluster model to the near critical regime \cite{DCM20}.

Before diving into the details, we briefly show how Theorem \ref{main} follows (as explained in \cite{cjn20}) from Theorem \ref{prop3}.

\begin{proof}[Proof of Theorem \ref{main}]
Let $H$ be such that $a =  H^{\frac{8}{15}}$ and $h = 1$. 
Let $\langle \sigma_0 ; \sigma_x \rangle  =  \langle \sigma_0  \sigma_x \rangle_{a,h} - \langle \sigma_0 \rangle_{a,h} \langle \sigma_x  \rangle_{a,h}$. 
Then from Theorem \ref{prop3}
 \begin{align*}
\langle \sigma_0 ; \sigma_x \rangle_{ H^{\frac{8}{15}}, 1} \leq C(1)   H^{\frac{2}{15}} \abs{x}^{-\frac{1}{4}} e^{-m(1) \abs{x}} 
\end{align*}
for $x \in H^{\frac{8}{15}}\Z^2$.  Using the relation $\langle \sigma_0 ; \sigma_x \rangle_{ H^{\frac{8}{15}}, 1} = \langle \sigma_{0} ; \sigma_{x'} \rangle_{1, H}$ whenever $x' = \frac{x}{H^{\frac{8}{15}}}$ we obtain
 \begin{align*}
\langle \sigma_0 ; \sigma_{x'} \rangle_{1, H} \leq C(1)  \abs{x'}^{-\frac{1}{4}} e^{-m H^{\frac{8}{15}} \abs{x'}} 
\end{align*}
for $x' \in \Z^2$.  Rescaling back to $a\Z^2$ yields the result. \end{proof}

In \cite{cjn20} a converse inequality is also proved using reflection positivity. A more probabilistic proof of the lower bound was given in \cite{CJNup}. We note that this shows that the correlation length is finite,  the mass gap exists and that critical exponent of the correlation length equals $\frac{8}{15} $. Further, as it is explained in \cite{cjn20} the exponential decay proven in Theorem \ref{main} directly translates into the scaling limit. 

\noindent Indeed, as in \cite{cjn20} if $\Phi^{a,h}$ is the near critical magnetization field given by 
\begin{align*}
\Phi^{a,h} = a^{ \frac{15}{8}} \sum_{x \in a \mathbb{Z}^2} \sigma_x \delta_x 
\end{align*} 
with $\{\sigma_x \}_{x \in a \mathbb{Z}^2} \in \{0,1\}^{a\Z^2} $,  it was proven in Theorem 1.4 of \cite{camia2016planar}  that $ \Phi^{a,h} $ converges in law to a continuum (generalized) random field $\Phi^h$. Let $C_0^\infty(\R^2)$ denote the set of smooth functions with compact support and let $\Phi^h(f)$ be $\Phi^h$ paired against $f \in C_0^\infty(\R^2)$. Then as in  \cite{cjn20} it holds that 
\begin{corollary}
Let $f,g \in C_0^\infty(\R^2)$, then there are $B_0, C_0 \in (0, \infty)$ such that 
\begin{align*}
\abs{\text{Cov}( \Phi^h(f), \Phi^h(g))}
 \leq	C_0 \int \int_{\R^2 \times \R^2} \abs{f(x)}\abs{g(x)} \abs{x-y}^{- \frac{1}{4}}  e^{- B_0 h^{\frac{8}{15} }\abs{x-y} }dx dy. 
\end{align*} 
\end{corollary}

 Starting with \cite{Zam89}, there has in the physics community been much interest in the masses of the Ising model \cite{Mus10} including possible connections to the exceptional Lie Algebra $E_8$ \cite{BG11} which has been investigated also experimentally \cite{C+10, E820, zhangexp} and numerically \cite{CH00}.  On the mathematical side, exponential decay was first rigorously proven in \cite{LP68} and in \cite{FR12} a linear upper bound for the mass was proven. Proving the correct scaling exponent is a further step towards rigorous results in this direction. For further rigorous developments see also \cite{CJNgau}.

\section{Preliminaries}
We start by briefly introducing the Ising model and its random cluster and random current representations that we will use to prove the result. 
Let $G = (V,E)$ be a finite graph. Then for each spin configuration $\sigma \in \{ \pm 1 \}^V$ and $h \geq 0$ define the energy
\begin{align*}
    H(\sigma) = - \sum_{xy \in E} \sigma_x \sigma_y  - h \sum_{x \in V } \sigma_x,
\end{align*}
where $h$ describes the effect of an external magnetic field. For each $A \subset V$ we let $\sigma_A = \prod_{x \in A} \sigma_x $ and define the correlation function as 
\begin{align*}
    \langle \sigma_A \rangle = \frac{\sum_{\sigma  \in \{ \pm 1 \}^V} \sigma_A \exp(- \beta H(\sigma)) }{Z}
\end{align*}
where $Z =\sum_{\sigma  \in \{ \pm 1 \}^V} \exp(- \beta H(\sigma)) $ is the partition function. In what follows, we will be concerned with the Ising model on the graph $a\Z^2$ which is obtained by taking the thermodynamics limit of finite graphs. For discussions about the thermodynamic limit we refer the reader to \cite{FV}. 

In both representations we implement the magnetic field using Griffiths' ghost vertex $\mathrm{g}$. This means that we consider the graph $G_{\text{ghost}} = (V \cup \{\mathrm{g} \}, E \cup E_\mathrm{g}) $ where $E_\mathrm{g} = \cup_{v \in V} \{ e_{v\mathrm{g}} \} $ are additional edges from every original vertex $v$ to the ghost vertex $\mathrm{g}$ (see for example \cite{DC17}). We will refer to the edges $E$ as \emph{internal} edges and to the edges $E_\mathrm{g}$ as \emph{ghost} edges. 

\subsubsection*{The random current representation}
Let us now introduce the random current representation which is a very effective tool in the study of the Ising model \cite{Raoufi2017,Aiz82, AD19, ADTW19,ABF87, ADS13,Sakai2005LaceEF,Shlosman1986SignsOT}. Further information can be found in \cite{DC17} and \cite{DC16}. The central building blocks in the random current representation of the Ising model on a graph $G_{\text{ghost}} = (V \cup \{\mathrm{g} \}, E \cup E_\mathrm{g}) $ are the currents $\nn \in {\N_0}^{E \cup E_\mathrm{g}}$. 
For each current $\nn$ we can define its sources $\partial \nn$ as the $v \in V$ where $ \sum_{xv \in E \cup E_g} \nn_{xv} $ is odd. Let further the \emph{weight} of each current $\nn$ be given by 
 \begin{align*}
 w(\nn) = \prod_{xy \in E} \frac{\beta^{\nn_{xy}}}{\nn_{xy}!} \prod_{x\mathrm{g}\in E_\mathrm{g}} \frac{(\beta h)^{\nn_{x\mathrm{g}}}}{\nn_{x\mathrm{g}}!}.
 \end{align*}
  A simple identity which connects the random currents to the Ising model is given as (2.4a) in \cite{ABF87}
\begin{align} \label{ghostcor}
\langle \sigma_{0} \sigma_{x} \rangle  =  \frac{ \sum_{\partial \nn = \{0,x\} } w(\nn) }{\sum_{\partial \nn = \emptyset} w(\nn)}.
\end{align}

\noindent Further, given a current $\nn$  define the traced current 
$\hat{\nn} \in \{0,1\}^{E \cup E_\mathrm{g}}$ by $\hat{\nn}(e) = 0$ if $\nn(e) = 0$ and $\hat{\nn}(e) = 1$ if $\nn(e) > 0$. Then $\pr_G^{A}$, the random current measure with sources  $A \subset V$, is the probability measure that satisfies
$
\pr_G^{A}(\nn) \propto w(\nn) 1_{\{ \partial \nn = A \} }. 
$
If $A$ and $B$ are either vertices in or subsets of $V \cup \{g \}$ we denote the event that they are connected in a configuration $\omega \in  \{0,1\}^{E \cup E_\mathrm{g}}$ by $A \cc B$, meaning that one vertex of $A$ is connected to one vertex of $B$. 

The traced random current measure  $\hat \pr_G^{A}$ gives each $\omega \in  \{0,1\}^{E \cup E_g}$ the probability
 \begin{align*}
 \hat \pr_G^{A}(\omega) =  \sum_{\partial \nn = A} \pr_G^{A}(\nn) 1_ { \{ \hat \nn = \omega \} }. 
 \end{align*}   
To ease the notation in what follows define  $\hat \pr_G^{\{0,x\}} \otimes \hat \pr_G^{\emptyset} $  to be the probability measure which assigns each $\omega \in \{0,1\}^{E \cup E_\mathrm{g}} $ the probability
\begin{align*} \hat \pr_G^{\{0,x\}} \otimes \hat \pr_G^{\emptyset} (\omega) =  \frac{1}{Z_\emptyset Z_{\{0,x\}}} \sum_{\partial \nn = \{0,x \}, \partial \mm = \emptyset} w(\nn) w(\mm) 1 \lbrack \widehat{\nn+\mm} = \omega \rbrack .
\end{align*}

\noindent   On the square lattice $a \Z^2$ in a magnetic field $a^{15/8}\, h$ we denote the non traced and traced single current measures by $\pr_{a,h}^{A}$ and $ \hat \pr_{a,h}^{A}$ respectively. The main part of what follows proves exponential decay of truncated correlations, but first we obtain  the correct front factor $a^\frac{1}{4}$. We do a similar trick as in \cite{cjn20} where we set the magnetic field $h$ to $0$ in the boxes of radius 1 around $0$ and $x$ and call that magnetic field $\vec{h}$.

\begin{proposition}\label{randomcurrentcor}
We have 
\begin{align*} 
\langle \sigma_0 ; \sigma_x \rangle_{a,h}  \leq \langle \sigma_0 ; \sigma_x \rangle_{a,\vec{h}}   = \langle \sigma_0 \sigma_x \rangle_{a,\vec{h}} \cdot  \hat \pr^{\{0,x \}}_{a,\vec{h}} \otimes  \hat \pr^{\emptyset}_{a,\vec{h}}( 0 \not \cc \mathrm{g})  \leq C a^{\frac{1}{4}} \hat \pr^{\{0,x \}}_{a,\vec{h}}( 0 \not \cc \mathrm{g}). 
\end{align*} 
\end{proposition}

\begin{proof}
The first inequality follows from the GHS inequality \cite{GHS70}. The second step used the switching lemma \cite{GHS70}  and  (\ref{ghostcor}).
Since the event $\{ 0 \not \cc \mathrm{g} \} $ is decreasing  the probability increases when $\hat \pr^{\emptyset}_{a, \vec{h}}$ is removed. Then the last inequality is a standard application of equation (1-arm) below (see also \cite{DHN11}). 
\end{proof}

\subsubsection*{The random cluster model}
Each configuration $\omega \in \{0,1\}^{E \cup E_\mathrm{g}}$ corresponds to a (spanning) subgraph of $G_{ghost}$. For each $e \in E \cup E_\mathrm{g}$ if $w_e = 1$ we say that $e$ is \emph{open} and if $w_e = 0$ we say that $e$ is \emph{closed}. There is a natural partial order $ \preceq $ on the configurations where  $ \omega \preceq  \omega' $ if $\omega'$ can be obtained from $ \omega$ by opening edges. An event $\mathcal{A}$ is \emph{increasing} if for any $\omega \in  \mathcal{A}$ it holds that $\omega \preceq \omega'$ implies $\omega' \in \mathcal{A}$. Let further  $k(\omega)$ be the number of clusters of vertices of the configuration $\omega$. 

The random cluster model with \emph{free boundary conditions}  $\phi_{G}^0$ is a percolation measure on the finite graph $G_{ghost} =(V \cup \{\mathrm{g} \},E \cup E_\mathrm{g})$ such that for every $\omega \in \{0,1\}^{E\cup E_\mathrm{g}}$   
\begin{align*}
    \phi_{G}^0({\omega}) \propto   2^{k(\omega)} \prod_{e \in E \cup E_\mathrm{g}} \frac{p_e}{1-p_e} 
\end{align*} 
where
 $p_e = 1_{ \{ e \text{ is internal and open} \} }   (1- \exp( -2 \beta)) + 1_{ \{ e  \text{ is ghost edge and open} \} }  (1- \exp( -2 \beta h))  + \frac{1}{2} 1_{ \{e\text{ closed}\}}$. In what follows, we will consider the \emph{free} random cluster model on some finite subsets $\Lambda$ of $a \Z^2$ and we will denote that measure by $ \phi_{\Lambda}^{a} $ at the same time fixing $\beta = \beta_c = \frac{ \log( 1+ \sqrt{2}) }{2} $ . Let $\Lambda_k(x)$ denote  the box of with side length $k$ around some point $x \in a \Z^2$ and let $\Lambda_k = \Lambda_k(0)$. Notice that $\Lambda_k$ only depends on the distance in $\R^2$ which is not affected when $a$ changes.   Further, let $A_{n,m}(x) = \Lambda_m(x) \slash  \Lambda_n(x)$ be the $(n,m)$ annulus around $x$ and $A_{n,m} = A_{n,m}(0)$. 
The random cluster model has many nice properties that we will use in what follows. 
\noindent Since the boundary conditions are free the random cluster model has stochastic domination in terms of the domain. This means that if $\Lambda_1 \subset \Lambda_2$ then for any increasing event $\mathcal{A}$, 
\begin{align}
\tag{MON}
 \phi_{\Lambda_1}^{a}( \mathcal{A}) \leq  \phi_{\Lambda_2}^{a}( \mathcal{A}).
\end{align}
Further, the (FKG)-inequality  \cite[Theorem 1.6]{DC17} states that for increasing events $\mathcal{A}, \mathcal{B}$ then
\begin{align}
\tag{FKG}
 \phi_{\Lambda}^{a}( \mathcal{A} \cap \mathcal{B}) \geq  \phi_{\Lambda}^{a}( \mathcal{A})  \phi_{\Lambda}^{a}( \mathcal{B}). 
\end{align}

\noindent We note that the 1-arm exponent for the random cluster model  \cite[Lemma 5.4]{DHN11} is given by
\begin{align}
\tag{1-arm}
C_1 a^{\frac{1}{8}}   \leq \phi_{\Lambda_1}^{a}(  0 \cc \partial \Lambda_1 )  \leq C_2 a^{\frac{1}{8}}. 
\end{align}

\noindent The following result was proven in \cite{cjn20} and it will also prove useful for us.  

\begin{lemma} (\cite{cjn20}, Proposition 1)  \label{ghosted}
Suppose that configuration of internal edges $\omega $ has clusters $C_1 , \dots, C_n$. Then 
\begin{align*}
\Prb^a_{\Lambda_{3}}( C_i \leftrightarrow \mathrm{g} \vert \omega ) = \tanh(h a^{\frac{15}{8}} \abs{C_i})
\end{align*}
and the events $\{ C_i \leftrightarrow \mathrm{g} \}$ given $\omega$  are mutually independent. 
\end{lemma}

\noindent Finally, we state a connection between the random currents with sources and the random cluster model.

\begin{theorem} (\cite{ADTW19}, Theorem 3.2)   \label{sourcecop}
Let $\{ X(e) \}_{e \in E}$ be independent Bernoulli percolation with parameter $(1- \exp(-\beta_e))$ with $\beta_e = \beta $ for $e \in E$ and $ \beta_e = \beta h $ for $e \in E_\mathrm{g}$.  Then define for each $e \in E \cup E_\mathrm{g}$ the configuration 
\begin{align*}
\omega(e) = \max\{\hat{\nn}(e), X(e) \}.
\end{align*} 
where $\hat{\nn} $ has the law of  ${\hat  \pr}^{\{ x,y \}}$ the traced random current with sources $\partial \nn = \{x,y \}$. Then 
$
\omega $ has the law of $  \phi_G^0( \cdot \mid x \cc y) $
which is  the random cluster measure conditioned on $\{x \cc y \}$. Hence, if $\mathcal{A}$ is a decreasing event then 
 \begin{align*}
 {\hat  \pr}^{\{ x,y \}}(\mathcal{A}) \geq \phi_G^0( \mathcal{A} \mid x \cc y). 
 \end{align*}
\end{theorem}



\noindent A key part in our result is the backbone exploration which we turn to next.

 \begin{figure}
\floatbox[{\capbeside\thisfloatsetup{capbesideposition={left,top},capbesidewidth=4cm}}]{figure}[\FBwidth]
{  \caption{The situation in the backbone exploration with the incoming edge $e_{u_{i-1}, u_i}$ coloured blue. The vertices $v_R, v_L, v_S$ are respectively to the right, left and straight of $u_i$ with respect to the incoming edge. As usual $\mathrm{g}$ denotes the ghost. }
    \label{incoming_edge} }
     { \includegraphics[scale =0.4]{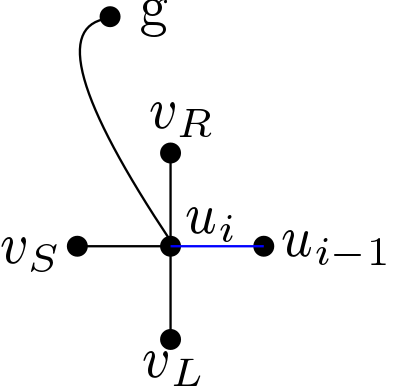}}
\end{figure}

\subsubsection*{Backbone exploration}
Let us first define the partially explored backbone. Suppose that $\nn$ is some current with sources $ \partial \nn = \{x,y\}$. Then there is a path between $x$ and $y$ with $\nn(e)$ odd for all edges $e$ along the path. The backbone is an algorithmic way of step-by-step constructing such a path until it hits some set of vertices $A \supset \{\mathrm{g}, y\}$  \footnote{ In the constructions in the litterature the set corresponding to $A$ usually does not necessarily contain the ghost, but for our purpose in this paper we include it. }. 
To do that, we define the sets of (explored) edges $ \emptyset = S_0 \subset S_1 \dots $ inductively. For each $i \geq 0$ the set $S_i$ is defined in such a way that $\nn$ restricted to $S_i$ has sources $\{x \} \triangle \{u_i \} $ for some vertex $u_i$. We will say that the backbones path up to step $i$ is $x = u_0, u_1, \dots u_i$.   If $u_i \in A$, let $S_{i+1} = S_i$ (and hence $S_{k} = S_i$ for all $k \geq i$).

If $u_i \not \in A$ we continue as follows. 
If $i = 0$ we consider the five edges incident to $u_0 = x$. Order them as $e_0, e_1, \dots, e_4$ with $e_0 = e_{x \mathrm{g}} $ and the other edges in arbitrary order. Since $x$ is a source of $\nn$ there is at least one $i$ such that $\nn(e_i) $ is odd. Let $k$ be the least such $i$ and let $S_1 =  \{ e_0, \dots e_k \}$. Then $u_1$ is such that $e_k = e_{u_0 u_1}$. In words, the backbone explored the edges $  \{ e_0, \dots e_k \}$ and walked to the vertex $u_1$.

For $i \geq 1$ we call the edge $e_{u_{i-1}, u_i}$ the \emph{incoming} edge to the vertex $u_i$. We can define an order on the remaining edges such that $(e_0, e_1, e_2, e_3) = (e_{u_i \mathrm{g}}, e_{u_i v_R}, e_{u_i v_L},  e_{u_i v_S})$ where $ e_{u_i v_R}, e_{u_i v_L},  e_{u_i v_S}$ denotes the edges that are right, left and straight with respect to the incoming edge. See also Figure \ref{incoming_edge}. 

Now, let $k$ be the least $i$ such that $\nn(e_i) $ is odd and $e_i \not \in S_i$. Notice that since $u_i \not \in A$ and $ \partial \nn = \{x,y \}$ there is always at least one such $i$. Define $S_{i+1} = S_i \cup \{ e_0, \dots e_k \}$. Then $u_{i+1}$ is such that $e_k = e_{u_i, u_{i+1}}$.
In words, the backbone walks on edges $e$ of odd $\nn(e)$ exploring in each step first the edge to the ghost and after that the edges right, left and straight with respect to the incoming edge in that order. The backbone path is the path of explored vertices $x = u_0, u_1, \dots $ and the explored backbone in step $i$ is $S_i$.


This sequence $\{ S_i \}_{i \in \N}$ stabilizes after a finite number of steps and we call the terminating set the \emph{backbone starting from $x$ explored up to $A$} and denote it by  $\bar \gamma_{x,A}(\nn)$. 
There is a path from $x$ to $A$ along the vertices $x= u_0, u_1, \dots, u_{\text{end}} $ with $u_{\text{end}}  \in A$  such that every edge $e$ in the path obeys that $e \in \bar \gamma_{x,A}(\nn)$ and has $\nn(e)$ odd. We call this path $ \gamma_{x,A}(\nn)$\footnote{Note that if $h =0$ then $ \bar \gamma_{x,\{\mathrm{g},y\}}(\nn)$ explores some edges in and around the path $\gamma_{x,\{\mathrm{g},y\}}(\nn)$  from $x$ to $y$ where $\nn(e)$ is odd for all traversed edges.}. The vertex $u_{\text{end}} $ we call  $\gamma_{x,A}^{ \text{end}}(\nn)$. If $u_{ \text{end} }$ is the ghost $\mathrm{g}$ we say that \emph{the backbone hits the ghost}.  

In what follows, we will work with events of the  $ \mathcal{Q} = \{ F = \bar \gamma_{0, \Gamma}(\nn)  \} $ where $\Gamma \supset \{\mathrm{g}, x \}$ is a set of vertices, and $F$ is a set of edges. Notice that by construction we can tell whether the explored backbone is $F$ only by looking at the edges in $F$ which means that $1_ \mathcal{Q} (\nn) = 1_ \mathcal{Q} ( \nn_F) $ where $ \nn_F$ is the current restricted to the set $F$. 

\noindent The partial backbone exploration is useful because of the following Markov property.

  \begin{figure}
  \centering
  \includegraphics[scale =0.065]{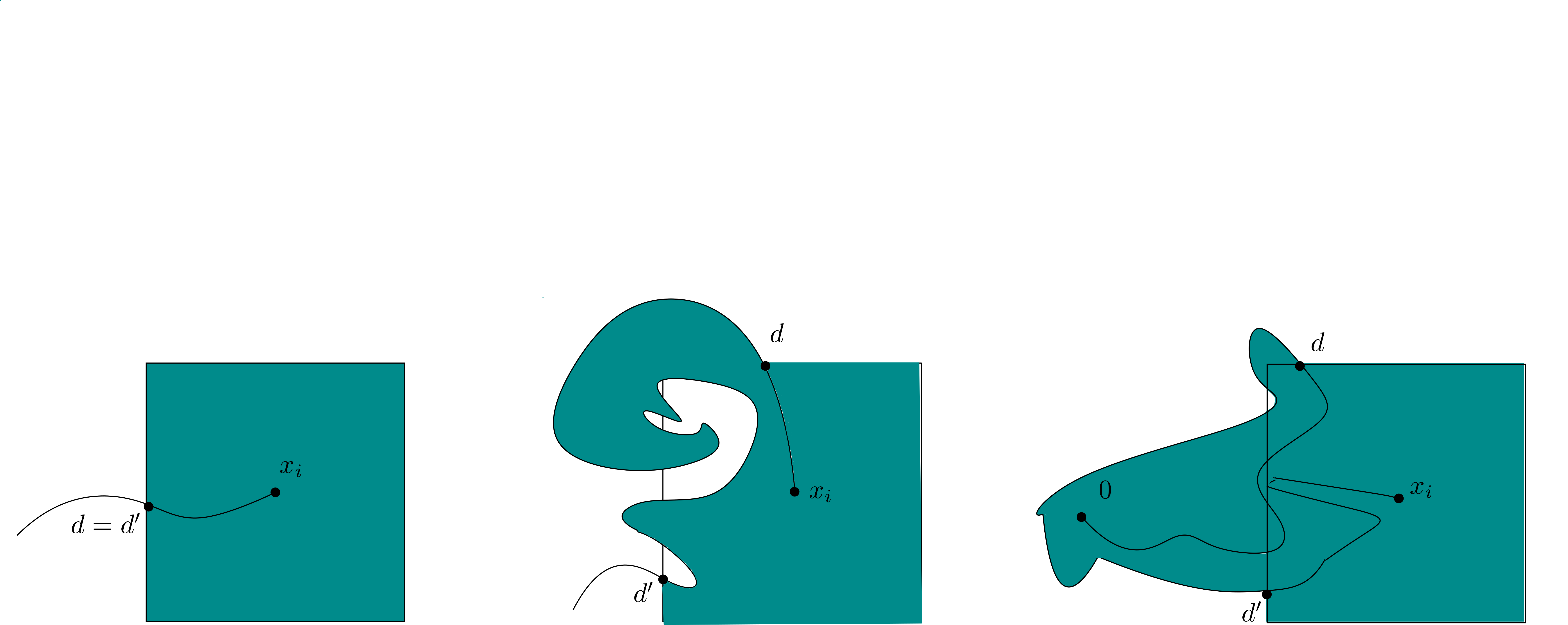}
    \caption{The explored backbone sketched in black together with the points $d$ and $d'$ in three different cases. The domain $D_i$ coloured green. Note that in the first case $d = d'$ and that in the last picture the position of $0$ is not to scale.  \label{domains} }
  \label{Domains}
\end{figure}

\begin{theorem} \label{backbone exploration}
Let $\Gamma \supset \{\mathrm{g}, x \}$ be a set of vertices, $F$ be a set of edges, $ \mathcal{Q} = \{ F = \bar \gamma_{0, \Gamma}(\nn)  \} $ and on the event $ \mathcal{Q} $ let $\tilde x =  \gamma_{0, \Gamma}^{\textit{end}}(\nn)  $ be the unique vertex in $\Gamma$ connected to $x$ in $\gamma_{0, \Gamma}(\nn) $.  Let $\mathcal{A}$ be an event such that $1_\mathcal{A}(\nn_{\Lambda}) 1_ \mathcal{Q} (\nn_{\Lambda}) = 1_\mathcal{A}(\nn_{\Lambda \slash F} ) 1_ \mathcal{Q} ( \nn_F) $. Then whenever $ \pr^{\{0,x\}}_{\Lambda}(  \mathcal{Q}  ) > 0 $ it holds that
\begin{align*}
\pr^{\{0,x\}}_{\Lambda}( \mathcal{A} \mid  \mathcal{Q} ) = \pr^{ \{ \tilde x, x \} }_{\Lambda \backslash F}(\mathcal{A}). 
\end{align*}
\end{theorem}
\begin{proof}
That $\nn \in \mathcal{Q} $ means that the explored backbone of $\nn$ up to $\Gamma$ is $F$. Thus, $F$ is the terminating set of the sequence $\{S_i\}_{i \in \N}$. Thus, on the event $ \mathcal{Q} $ the current $\nn$ restricted to $F$ must have sources $ \partial  \nn_{F} = \{0\} \triangle \{u_\text{end} \} = \{0,\tilde x \}$. Since $\nn = \nn_{\Lambda \backslash F}+\nn_{F} $ it holds that 
\begin{align*}
\{0,x \} = \partial \nn  =  \partial  \nn_{\Lambda \backslash F} \triangle \partial  \nn_{F} =  \partial  \nn_{\Lambda \backslash F} \triangle \{0, \tilde x \}.
\end{align*}
 So for $\nn \in \mathcal{Q} $ then $\partial \nn_{\Lambda \backslash F}  = \{ \tilde x, x \}. $
The map $\nn \mapsto (\nn_F, \nn_{\Lambda \backslash F})$ is a bijection from $ \left \{\nn \in \mathcal{Q} \mid \partial \nn = \{0,x \} \right \}$ to $\left \{ (\nn_F, \nn_{\Lambda \backslash F}) \mid  \nn_F + \nn_{\Lambda \backslash F} \in \mathcal{Q},  \partial \nn_F = \{0, \tilde x\}, \partial \nn_{\Lambda \backslash F} = \{\tilde x, x \} \right \}$ with inverse $(\nn_F, \nn_{\Lambda \backslash F}) \mapsto \nn_F + \nn_{\Lambda \backslash F}$. 
Thus, for any function $f: {\N_0}^{E \cup E_\mathrm{g}} \to \R$ it holds that 
\begin{align*}
\sum_{\partial \nn = \{0,x \}} f( \nn) 1_ \mathcal{Q} (\nn) = \sum_{ \substack{ \partial \nn_F = \{0, \tilde x\} \\ \partial \nn_{\Lambda \backslash F} = \{\tilde x, x \}} } f( \nn) 1_ \mathcal{Q} (\nn).
\end{align*}
Since $w(\nn_F + \nn_{\Lambda \backslash F}) = w(\nn_F) \cdot w(\nn_{\Lambda \backslash F})$  and the fact that $ \mathcal{Q} $ only depends on edges in $F$, 
the double sum below factorizes and 
\begin{align*}
\pr^{\{0,x\}}_{\Lambda}( \mathcal{A} \mid  \mathcal{Q} )
& = \frac{ \sum_{\partial \nn = \{0,x \}} w(\nn) 1_\mathcal{A}(\nn)  1_ \mathcal{Q} (\nn) }{ \sum_{\partial \nn = \{0,x \} } w(\nn) 1_ \mathcal{Q} (\nn)} \\
 &= \frac{ \sum_{\partial \nn_F= \{0, \tilde x\}, \partial \nn_{\Lambda \backslash F} = \{\tilde x, x \} } w(\nn_F + \nn_{\Lambda \backslash F}) 1_\mathcal{A}(\nn) 1_ \mathcal{Q} (\nn) }{\sum_{\partial \nn_F = \{0, \tilde x\}, \partial \nn_{\Lambda \backslash F} = \{\tilde x, x \} }w(\nn_F+ \nn_{\Lambda \backslash F}) 1_ \mathcal{Q} (\nn)} \\
 &= \frac{ \sum_{\partial \nn_F= \{0, \tilde x\}} w(\nn_F) 1_ \mathcal{Q} (\nn_F)  \sum_{ \partial \nn_{\Lambda \backslash F} = \{\tilde x, x \} } w(\nn_{\Lambda \backslash F}) 1_\mathcal{A}(\nn_{\Lambda \slash F} )   }{\sum_{\partial \nn_F = \{0, \tilde x\}}  w(\nn_F) 1_ \mathcal{Q} (\nn_F)  \sum_{ \partial \nn_{\Lambda \backslash F} = \{\tilde x, x \} } w( \nn_{\Lambda \backslash F}) } \\
& =  \frac{ \sum_{ \partial \nn_{\Lambda \backslash F} = \{\tilde x, x \} } w(\nn_{\Lambda \backslash F}) 1_\mathcal{A}(\nn_{\Lambda \backslash F}) }{ \sum_{ \partial \nn_{\Lambda \backslash F} = \{\tilde x, x \}} w(\nn_{\Lambda \backslash F}) }  = \pr^{ \{ \tilde x, x \} }_{\Lambda \backslash F}(\mathcal{A}).
\end{align*}
\end{proof}

\begin{figure}
\floatbox[{\capbeside\thisfloatsetup{capbesideposition={left,top},capbesidewidth=4cm}}]{figure}[\FBwidth]
{\caption{Sketch the order $\preceq$ on $\partial \Lambda_1(x_i)$ used when defining $D_i$. Two points $a,b \in \partial \Lambda_1(x_i)$ have $a \preceq b$ if the number of the segment of $a$ is smaller than the number of the segment of $b$ or if $a$ and $b$ are inside the same segment and $a$ is earlier than $b$ with respect to the arrow on that segment. \label{clock}}}
{\includegraphics[width=5cm]{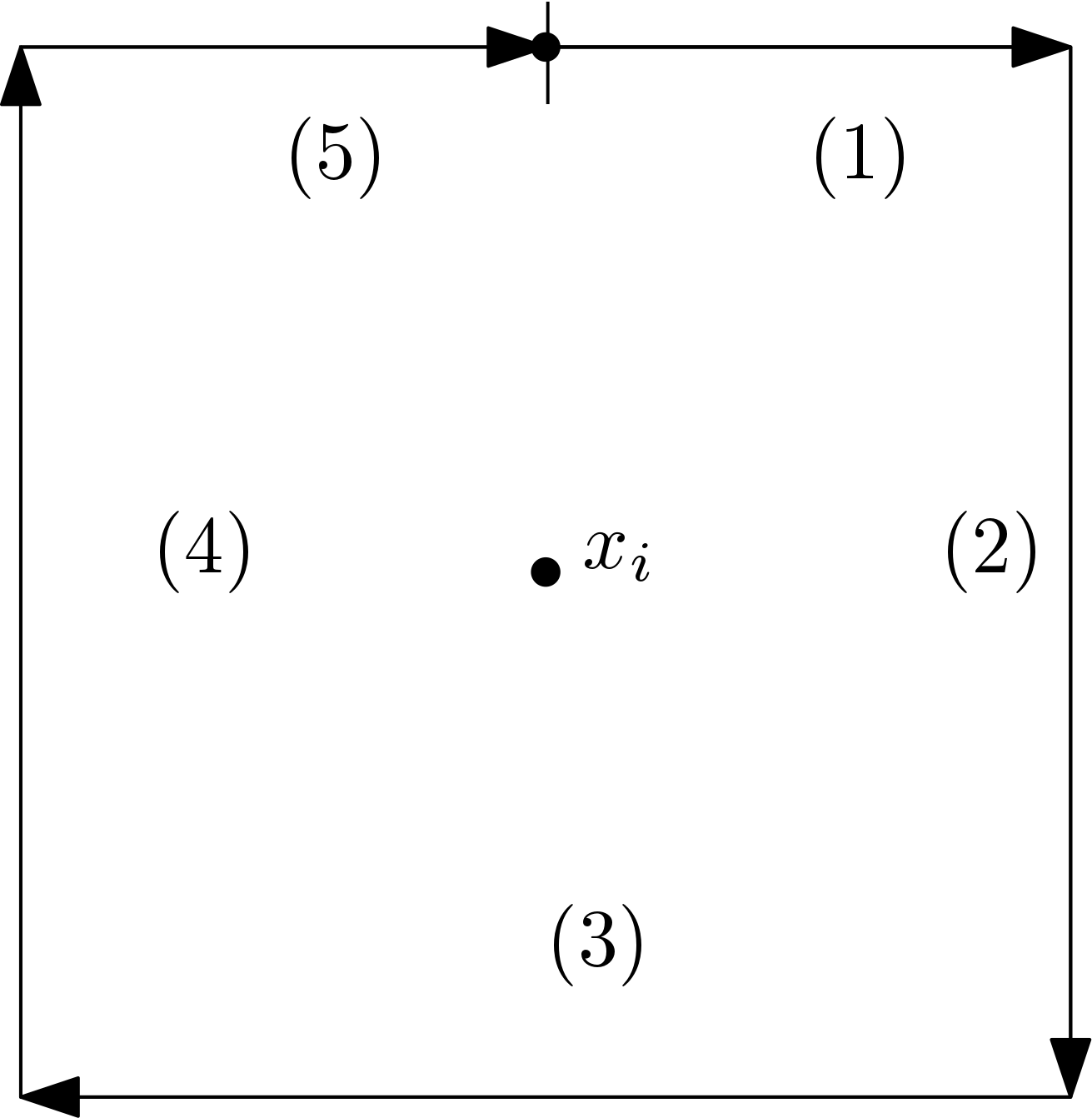}}
\end{figure}

\section{Main result}
We now prove Theorem \ref{prop3} given a result that we then prove later. 

Before starting the proof, we go through some notation that we will use throughout the main section. Let $n$ be such that $x \in A_{9n, 9(n+1)}$. In what follows, we will consider the case $\abs{x} \geq 36 $ which means that $n \geq 4$. In the proof of Theorem \ref{prop3} we tie it together with the case $\abs{x} < 36$. We also let $\Lambda$ be any box which contains $ \Lambda_{9(n+1)}$. Everything we prove will be independent of this $\Lambda$.  Later we let $\Lambda \Uparrow \Z^2$. 

\begin{figure}
\floatbox[{\capbeside\thisfloatsetup{capbesideposition={left,top},capbesidewidth=4cm}}]{figure}[\FBwidth]
{\caption{Sketch of the overall approach.  Either the backbone goes through the ghost or we can define the $x_i$ as shown. Later, we show that it is probable that $x_i$ is connected to the ghost within each of the black squares.\label{overall}}}
{\includegraphics[width=5cm]{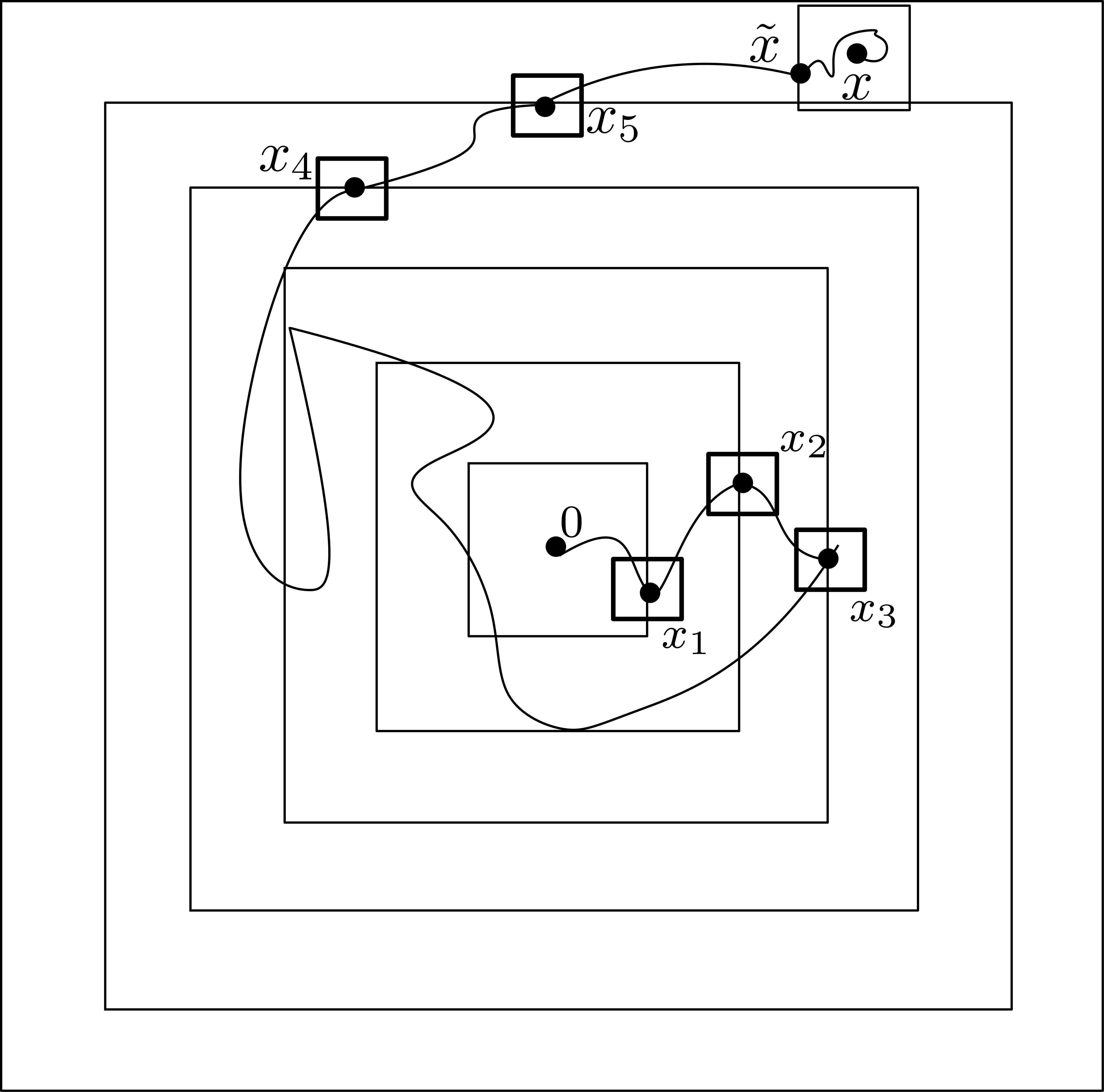}}
\end{figure}

We will explore the backbone partially in steps up to the annuli $A_{9i, 9(i+1)}$. Suppose that in this exploration the backbone does not hit the ghost $\mathrm{g}$, which we can assume in our application. Then define $x_i = \gamma_{0,  \Lambda_{9i}^c}^{\text{end}}(\nn)  \in A_{9i, 9(i+1)}$. Thus, $x_i$ is random variable corresponding to the first vertex the backbone hits in the $i$-th annulus of the form $A_{9i, 9(i+1)}$ see also Figure \ref{overall}. 

 Further, to ease notation we let $\bar \gamma_i = \bar  \gamma_{0,  \Lambda_{9i}^c}(\nn) $. Thus, $\bar \gamma_i$ is the set of edges explored until the backbone hits the $i$-th annulus. Similarly, to ease notation let $ \gamma_i =  \gamma_{0,  \Lambda_{9i}^c}(\nn) $. Thus, $\gamma_i$  is the path $u_0, \dots, u_{\text{end}} = x_i =  \gamma_{0,  \Lambda_{9i}^c}^{\text{end}}(\nn) $ explored until the backbone hits the $i$-th annulus.  Let  $\mathcal{Q}_{ \bar \gamma_{i}}$ be the event that the backbone explored is $\bar \gamma_i$. 

A technical detail is that to account for removing the magnetic field in the box of size 1 around $x$ we explore the backbone partially also from $x$  until it leaves the box of size $2$. Denote the explored backbone $\bar \gamma_{x, \Lambda_2(x)^c}(\nn)$ by $\bar \gamma_x$, name the first point hit outside $\Lambda_2(x)$ by $\tilde x = \gamma_{x, \Lambda_2(x)^c}^{\text{end}}(\nn)$  and let the event that the explored backbone is $\bar \gamma_x$ be denoted $\mathcal{Q}_{\bar \gamma_x}$.

The set $ \bar \gamma_i$ contains the path $\gamma_i$ from $0$ to $x_i$. Thus, $ \bar \gamma_i $ will intersect $\partial \Lambda_1(x_i)$ one or more times. Since $x_i$ is the first time the annulus $A_{9i,9(i+1)}$ is intersected  the set $ \bar \gamma_i \cap \partial \Lambda_1(x_i)$ is contained within one half of $\partial \Lambda_1(x_i)$. Let $d, d'$ denote the points in $ \bar \gamma_i \cap \partial \Lambda_1(x_i)$  that are most clockwise and anticlockwise with respect to some way of walking around $\partial \Lambda_1(x_i)$, see also Figure \ref{domains}.  

More formally, we consider an order $ \preceq $ of the points in $\partial \Lambda_1(x_i)$ and then define $d,d'$ to be the minimal and maximal element of $ \bar \gamma_i \cap  \partial \Lambda_1(x_i)$ with respect to this ordering. Let us define the order in the case where $x_i$ is in the right side of the annulus (which is the case $x_1, x_2, x_3$ on Figure \ref{overall}) generalising to the other cases is straightforward.  To do that, we split $\partial \Lambda_1(x_i)$ and define the order with respect to the segments and arrows as shown on Figure \ref{clock}.

Now, define $ \Omega_i = \{v \in \partial \Lambda_1(x_i) \mid d' \preceq v  \preceq d \} $ and let $ \tilde{\Omega}_i = \partial \Lambda_1(x_i) \backslash \Omega_i $.
Let $\Sigma_i$ be the graph obtained by removing $\bar \gamma_i \cup \bar \gamma_x$ from $\Lambda$.  
Then we can define the domain $D_i$ to be the connected component of $x_i$ in the  graph induced by the vertices of $\Sigma_i$ without $\tilde{\Omega}_i$. See also Figure \ref{domains}. In the following claim we show how our order of exploration with respect to the incoming edge implies that $D_i$ is a bounded domain. 

\begin{figure}
\floatbox[{\capbeside\thisfloatsetup{capbesideposition={left,top},capbesidewidth=8cm}}]{figure}[\FBwidth]
{\caption{Sketch of the case where the backbone explores the same vertex twice as in the proof of Claim \ref{free_bc}. Suppose that the backbone entered and left the vertex along edges of the same colour (orange or blue). Since we in the backbone exploration always explore turning to the right with respect to the incoming edge first the backbone path can never cross itself. Thus, a non-crossing  deformation as shown is always possible.  \label{no_crossing}}}
{\includegraphics[scale = 0.03]{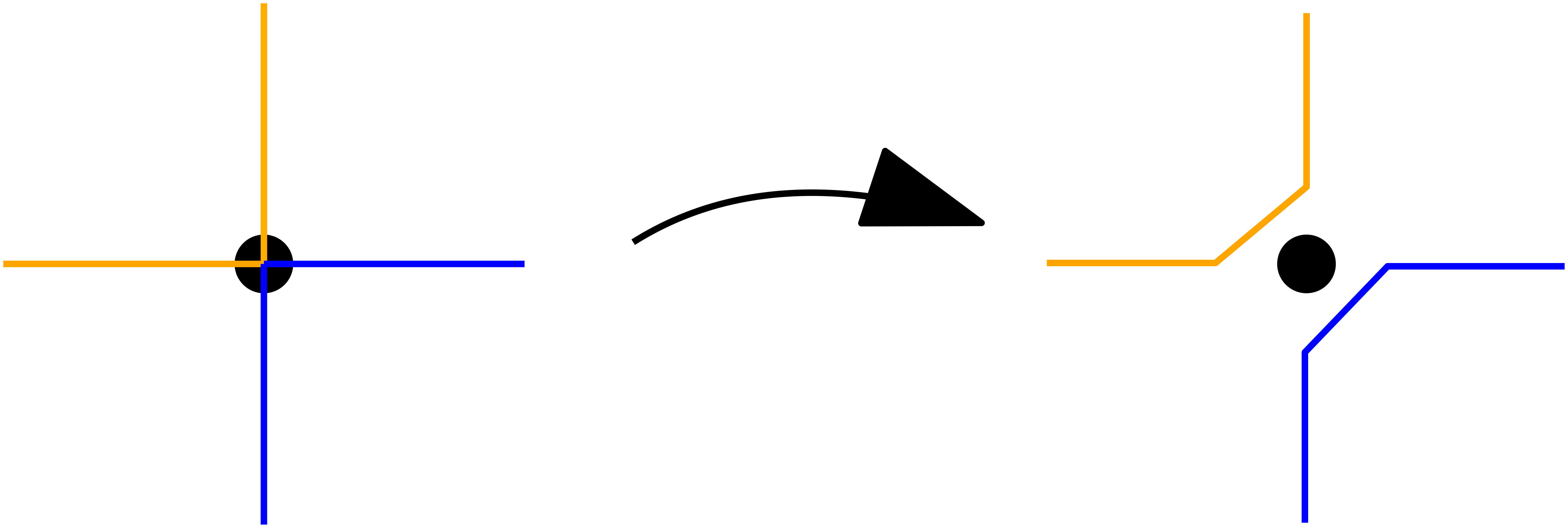}}
\end{figure}

\begin{claim} \label{free_bc} 
The set $D_i \subset \Lambda_{9(i+1)}$ and it only depends on the current $\nn_{\Lambda_{9i}}$.
\end{claim} 
\vspace{-0.7cm}
\begin{proof}
Since the vertex $d$ is explored by the backbone it is either on the backbone path  $\gamma_i$ in which case we let $v =d$. Otherwise, there is an edge $ e_{d v}$ from $d$ to a vertex on the backbone path that we call $v$. Similarly, we can define a vertex $v'$ taking $d'$ as the starting point. 
 Let $P_i$ be the subpath of $\gamma_i$ which goes either from $v$ to $v'$ or from $v'$ to $v$ and extended by the edges $ \{e_{d v} \}$ and/or $ \{e'_{d' v'} \} $ if $d, d'$ are not on the backbone path. By construction the path $P_i$ is edge self-avoiding, but  we do not know that $P_i$ is vertex self-avoiding and hence non self-intersecting. However, due to the way we explore the edges of the backbone with respect to the incoming edge  if there is a vertex which is hit by the backbone path twice (i.e. $u_i = u_j$ for some $i \neq j$) then the backbone path must turn $90^\circ$ twice at that vertex. This means that we can deform the path slightly to be non-intersecting (see Figure \ref{no_crossing}). 

Since $P_i$ is a path between $d$ and $d'$ and further $\tilde \Omega_i$ is also a path between $d$ and $d'$ along $\Lambda_1(x_i)$ that by definition of $d$ and $d'$, $P_i$ and  $\tilde \Omega_i$  do not intersect. Thus, if we glue them together then $P_i \cup \tilde \Omega_i $ is a closed non-intersecting path, which therefore encloses a domain $Q_i$.
 Now, assume for contradiction that $\delta: x_i \to Q_i^c$ is a path. Since we have removed all the vertices in $\tilde \Omega_i$ including $d$ and $d'$ it is impossible for $\delta$ to exit $Q_i$ through a vertex in $\tilde \Omega_i$. Since in the backbone exploration we remove the explored edges on the backbone path all remaining vertices have degree at most 2 in $\Sigma_i \backslash \tilde \Omega_i$. It is only possible to have a path exiting $Q_i$ if it crosses the path $P_i$ through a vertex  of degree 2.  The way we explore the backbone imply that if two edges remain they must have an angle of $90^\circ$. Therefore, the remaining edges do not cross $P_i$ and hence there is no such path $\delta$. Since $D_i$ is the connected component of $x_i$ it holds that 
 $D_i \subset Q_i$ and boundedness of $D_i$ follows from boundedness of $Q_i$. 
\end{proof}

 \noindent To finish the setup we finally define 
 \begin{align*}
 \mathcal{R}_i^* = \{\nn \mid \partial \Lambda_2(x_i)  \overset{A_{2,4}(x_i) \backslash (D_i \cup \bar \gamma_i)}{  \not \cc} \partial \Lambda_4(x_i) \}.
 \end{align*}
Notice that if $\hat{\nn}_1 = \hat{\nn}_2$ for two currents $\nn_1,  \nn_2$ and $\nn_1 \in  \mathcal{R}_i^* $ then  $\nn_2  \in  \mathcal{R}_i^*$. Thus, $ \mathcal{R}_i^*$  only depends on the traced and not on the full current. 
Define the corresponding connection event either for the traced current or for the random cluster measure by 
 \begin{align*}
 \hat{\mathcal{R}}_i^* =  \{ \partial \Lambda_2(x_i)  \overset{A_{2,4}(x_i) \backslash (D_i \cup \bar \gamma_i)}{  \not \cc} \partial \Lambda_4(x_i) \}  \subset \{0,1\}^{ A_{2,4}(x_i) \backslash (D_i \cup \bar \gamma_i)}. 
 \end{align*}
 Notice that 
$ \hat{\mathcal{R}}_i^* = 
\left \{ \hat{\nn} \mid \nn \in   \mathcal{R}_i^* \right \} $.  Hence, it holds that 
\begin{align} \label{fulltotraced} 
 \pr^{ \{x_i,  \tilde x \}}_{\Lambda \backslash (\bar \gamma_i \cup \bar \gamma_x),  \vec h} \left( \mathcal{R}_i^* \right) =  \hat{\pr}^{ \{x_i,  \tilde x \}}_{\Lambda \backslash (\bar \gamma_i \cup \bar \gamma_x),  \vec h} \left(\hat{ \mathcal{R}}_i^* \right).
\end{align} 

\noindent The following proposition will yield the main result given Proposition \ref{prop5} below. The idea of the proof is first to use the backbone exploration and then show that for every macroscopic step, with a strictly positive probability there is a connection to the ghost. 
To get the correct front factor $a^{\frac{1}{4}}$, we also partially explore the backbone also from the end around $x$ until $ \tilde x = \gamma_{x, \Lambda_2(x)^c}^{\text{end}}(\nn)$.

\begin{proposition} \label{mainthm}
Suppose that for all $\tilde x \in \partial \Lambda_2(x)$, all realisations of the backbone $\bar \gamma_x$ from $x$ to $\tilde x$, all $1 \leq i \leq n $ and realisations of the backbone from $0$ to $x_i$ denoted $\bar \gamma_i$ and $a \leq 1$ that
\begin{align*}
 \pr^{ \{x_i,  \tilde x \}}_{\Lambda \backslash (\bar \gamma_i \cup \bar \gamma_x),  \vec h} \left( \mathcal{R}_i^* \right) \geq c  
\end{align*}
 uniformly in any ($x$-dependent) $\Lambda$ sufficiently large. 
Then  
\begin{align*}
\langle \sigma_0 ; \sigma_x  \rangle_{a,h} \leq C a^{\frac{1}{4}}  \exp \left( - M(h) \abs{x} \right) 
\end{align*}
for $\abs{x} \geq 36$ and where $M(h)$ does not depend on $a$.  
\end{proposition}
\begin{proof}
Let $\mathcal{H}$ be the event that the backbone explored from $0$ hits $x$  ( i.e. does not hit the ghost).  
 Notice that $ \{ \tilde x = \mathrm{g}\}  \cap \mathcal{Q}_{\bar \gamma_x} \cap \mathcal{H} = \emptyset$ so when we condition $\mathcal{H}$ on all possible events $\mathcal{Q}_{\bar \gamma_x}$ we can omit those where $\tilde x = \mathrm{g}$.  In other words, if $ \tilde x = \mathrm{g}$ then the backbone explored from $0$ would necessarily hit the ghost since after the partial backbone exploration then $\mathrm{g}$ and $0$ would be the only two vertices with odd degree. 
 
 Further, 
$
1_{\mathcal{H}}(\nn_{\Lambda}) 1_ {\mathcal{Q}_{\bar \gamma_x}  } (\nn_{\Lambda}) = 1_{\mathcal{H}}(\nn_{\Lambda \backslash  \bar \gamma_x } ) 1_ {\mathcal{Q}_{\bar \gamma_x}  } ( \nn_{\bar \gamma_x})
$
 so by the backbone exploration Theorem \ref{backbone exploration} 
\begin{align}\label{heq} 
\pr^{\{0,x\}}_{\Lambda, \vec h}(\mathcal{H}) = \sum_{\bar \gamma_x} \pr^{\{0,x\}}_{\Lambda,\vec h}(\mathcal{H} \mid \mathcal{Q}_{\bar \gamma_x}) \pr^{\{0,x\}}_{\Lambda,\vec h}( \mathcal{Q}_{\bar \gamma_x} ) =  \sum_{\bar \gamma_x} \pr^{\{0, \tilde x\}}_{\Lambda \backslash \bar \gamma_x, \vec h}(\mathcal{H}) \pr^{\{0,x\}}_{\Lambda, \vec h}(\mathcal{Q}_{\bar \gamma_x})
\end{align} 
where we from now on assume that $\tilde x \neq \mathrm{g}$ which means that $\tilde x \in \partial \Lambda_2(x) $.

For each $1 \leq i \leq n$ let $\mathcal{G}_i$ be the event that the backbone hits the annulus $A_{9i, 9(i+1)}$ before hitting the ghost. Given a current configuration in $\mathcal{G}_i$ we know that the vertices $x_j$ exist for $1 \leq j \leq i $. 
Further,  if we define $\mathcal{G}_0 = \left \{ \nn \mid \partial \nn = \{0,x\} \right \} $ then $\mathcal{G}_{i+1} \subset \mathcal{G}_i$ for each $0 \leq i \leq n-1$ as well as $\{0 \not \cc \mathrm{g} \}  \subset \mathcal{H} \subset \mathcal{G}_n$. 
Since  $\pr^{\{0, \tilde x\}}_{\Lambda \backslash \bar \gamma_x, \vec h}(\mathcal{H}) \leq  \pr^{\{0, \tilde x\}}_{\Lambda \backslash \bar \gamma_x, \vec h}(\mathcal{G}_n)$ and bounding $\pr^{\{0, \tilde x\}}_{\Lambda \backslash \bar \gamma_x, \vec h}(\mathcal{H}) $ uniformly in $\bar \gamma_x$ bounds $ \pr^{\{0,x\}}_{\Lambda, \vec h} \left( \mathcal{H} \right)  $ through (\ref{heq}). 
It means that to bound $  \pr^{\{0,x\}}_{\Lambda, \vec h}( 0 \not \cc \mathrm{g} ) $  it suffices to  show exponential decay of $  \pr^{\{0, \tilde x\}}_{\Lambda \backslash \bar \gamma_x, \vec h}(\mathcal{G}_n)$ uniformly in $\bar \gamma_x$. 
Notice that given $ \mathcal{Q}_{ \bar \gamma_{i-1}} $ then  $\mathcal{G}_i$ depends only on edges in $\Lambda \backslash ( \bar \gamma_{i-1} \cup  \bar \gamma_x)$ so 
\begin{align*}
1_{\mathcal{G}_i}(\nn_{\Lambda \backslash \bar \gamma_x})
 1_ { \mathcal{Q}_{ \bar \gamma_{i-1}} } (\nn_{ \Lambda \backslash  \bar \gamma_x })
  = 1_{\mathcal{G}_i}(\nn_{\Lambda \backslash ( \bar \gamma_{i-1} \cup  \bar \gamma_x)} ) 1_ {\mathcal{Q}_{ \bar \gamma_{i-1}} } ( \nn_{(\bar \gamma_{i-1} \cup  \bar \gamma_x)})
\end{align*}
 and by the backbone exploration Theorem \ref{backbone exploration} 
\begin{align*}
    \pr_{\Lambda \backslash \bar \gamma_x, \vec h}^{\{0, \tilde x\}}(\mathcal{G}_n)  &=   \left \lbrack \prod_{i=1}^n   \pr_{\Lambda \backslash \bar \gamma_x, \vec h}^{\{0, \tilde x\}}(\mathcal{G}_i \mid \mathcal{G}_{i-1})  \right \rbrack  \pr_{\Lambda \backslash \bar \gamma_x, \vec h}^{\{0, \tilde x\}}(\mathcal{G}_0) \\
    &=  \prod_{i=1}^n \sum_{\mathcal{Q}_{ \bar \gamma_{i-1}} \in \mathcal{G}_{i-1}}   \pr_{\Lambda \backslash \bar \gamma_x, \vec h}^{\{0, \tilde x\}}(\mathcal{G}_i \mid \mathcal{Q}_{  \bar \gamma_{i-1}})   \pr_{\Lambda \backslash \bar \gamma_x, \vec h}^{\{0, \tilde x\}}(  \mathcal{Q}_{ \bar \gamma_{i-1}} \mid \mathcal{G}_{i-1})  \\
    & \leq \prod_{i=1}^n \sum_{\mathcal{Q}_{ \bar \gamma_{i-1}} \in \mathcal{G}_{i-1}} \pr_{\Lambda \backslash ( \bar \gamma_{i-1}  \cup  \bar \gamma_x), \vec h }^{\{x_{i-1}, \tilde x\}}(\mathcal{G}_i ) \pr_{\Lambda \backslash \bar \gamma_x, \vec h}^{\{0, \tilde x\}}( \mathcal{Q}_{ \bar \gamma_{i-1}} \mid \mathcal{G}_{i-1}) \leq (1-c)^{n-2} 
\end{align*}
where in the last step we used $ \pr_{\Lambda \backslash ( \bar \gamma_{i-1} \cup  \bar \gamma_x), \vec h }^{\{x_{i-1}, \tilde x\}}(\mathcal{G}_i ) \leq  \pr_{\Lambda \backslash ( \bar \gamma_{i-1} \cup  \bar \gamma_x) , \vec h}^{\{x_{i-1}, \tilde x \}}({\mathcal{R}^{*c}_{i-1}}) \leq 1-c$.  
Now, by the remarks in the beginning of the proof it follows from Proposition \ref{randomcurrentcor}  that
\begin{align*}
\langle \sigma_0 ; \sigma_x \rangle_{\Lambda,h,a}   \leq C a^{\frac{1}{4}} \pr^{\{0,x \}}_{\Lambda,\vec{h}} \left(  0 \not \cc \mathrm{g}  \right) \leq  C a^{\frac{1}{4}} (1-c)^{n-2}  =  C a^{\frac{1}{4}} \exp( - M(h) \abs{x}). 
\end{align*}
The inequality passes to the infinite volume limit since the constants are independent of $\Lambda$. 
\end{proof}

\noindent Next, we will move from the random current event $\mathcal{R}_i^*$ to the traced current or random cluster event $\hat{\mathcal{R}}_i^* $. In the remaining section we will prove the following proposition which only concerns the random cluster model. Here and in all of the following by $\geq c$ we mean larger than a (possibly different each time) strictly positive constant which is uniform in $a \leq 1$, $h$ and the explored backbones $\bar \gamma_i$ and $\bar \gamma_x$ .   

\begin{proposition} \label{prop5}
There is a $h_0 > 0 $ such that for all realisations of the backbone $ \bar \gamma_i$ up to $x_i$, all $\bar \gamma_x$ explored until $\tilde x \in \partial \Lambda_2(x)$ and all $a \leq 1, h < h_0$ then 
 \begin{align*}
\phi^{0,a}_{\Lambda \backslash ( \bar \gamma_i \cup \bar \gamma_x) , \vec h} \left( \hat{\mathcal{R}}_i^* \mid x_i \cc \tilde x \right) \geq c 
\end{align*}
 uniformly in any $\Lambda$ sufficiently large around $x$. 
\end{proposition}

\noindent Now, collecting the results we can prove the main theorem assuming Proposition \ref{prop5} which is in the language of the random cluster model where many more tools are available than for random currents most notably the RSW.
\begin{proof} [Proof of Theorem \ref{prop3}]
\noindent By (\ref{fulltotraced}), the monotone coupling from Theorem \ref{sourcecop} along with the fact that $ \hat{\mathcal{R}}_i^*$  is a decreasing event and Proposition \ref{prop5} we get that
\begin{align*}
\pr^{ \{x_i,  \tilde x \}}_{\Lambda \backslash (\bar \gamma_i \cup \bar \gamma_x),  \vec h} \left( \mathcal{R}_i^* \right) = \pr^{ \{x_i, \tilde x \}}_{\Lambda \backslash (\bar \gamma_i \cup \bar \gamma_x),  \vec h}( \hat{\mathcal{R}}_i^*) \geq \phi^{0,a}_{\Lambda \backslash (\bar \gamma_i \cup \bar \gamma_x) , \vec h}\left(  \hat{\mathcal{R}}_i^*  \mid x_i \cc \tilde x  \right) \geq c. 
\end{align*} 
 Thus, we can apply Proposition \ref{mainthm}.  In Proposition \ref{prop5} we only have the result for sufficiently small $h$, but this suffices by the GHS-inequality \cite{GHS70}. To account for the constraint $\abs{x} \geq 36$ in Proposition \ref{mainthm} and get the correct front factor notice that from the GHS inequality  and Proposition 5.5 in \cite{DHN11} some $B>0$ it holds for all $x \in a \Z^2$ that 
\begin{align*} 
\langle \sigma_0 ; \sigma_x \rangle_{a,h}  \leq \langle \sigma_0 ; \sigma_x \rangle_{a,0} \leq B \left( \frac{a}{\abs{x}} \right)^{\frac{1}{4}}. 
\end{align*}
Using that for $\abs{x} \geq K(h)$ it holds that 
\begin{align*}
\langle \sigma_0 ; \sigma_x  \rangle_{a,h} \leq C a^{\frac{1}{4}}  \exp \left( - \frac{M(h)}{2} \abs{x} \right) \exp \left( - \frac{M(h)}{2} \abs{x} \right)
\leq C \left(\frac{a}{\abs{x}} \right)^{\frac{1}{4}}  \exp \left( - \frac{M(h)}{2} \abs{x} \right). 
\end{align*}
By putting $C(h) = \max \{B, C\}  \exp \left( \frac{M(h)}{2} K(h) \right) $ and $m(h) = \frac{M(h)}{2} $ our main result Theorem \ref{prop3} follows. 

\end{proof}

\section{ Proof of Proposition \ref{prop5}} 

\noindent Recall that $\Sigma_i = \Lambda \backslash ( \bar \gamma_i \cup \bar \gamma_x)$. To ease the notation here and in what follows we define  $\phi_{\Sigma_i} = \phi^{0,a}_{\Sigma_i , \vec h} = \phi^{0,a}_{\Lambda \backslash ( \bar \gamma_i \cup \bar \gamma_x) , \vec h}$.  Further,  for a set $\Gamma$ let $x \cco{\Gamma^+} y$ denote the event that $x$ and $y$ are connected in  $\Gamma \cup \{\mathrm{g} \} $ and similarly by $x \cco{\Gamma} y$ that $x$ and $y$ are connected in $\Gamma$  \emph{not using the ghost}. Define the domain $T_i$ to be all points in $D_i$ as well as all points in $\Lambda_2(x_i)$ that can be reached from $x_i$ without using edges in $\bar \gamma_i$ or $\partial \Lambda_2(x_i)$. Further, define $ \{ x_i \cco{ T_i^+} \mathrm{g} \}$ to be the event that $x_i$ is connected within the domain $T_i$ to some vertex $v$ where the edge from $v$ to $\mathrm{g}$ is open. Define $\tilde D$ and $\tilde T$ similarly to  $D_i$ and $T_i$ with $\tilde x$ instead of $x_i$.
Define also $A_i = \partial \Lambda_2(x_i)  \cap T_i^c$ and $\tilde A = \partial \Lambda_2( \tilde x) \cap  \tilde T^c$.

\begin{proposition} \label{prop2}
 Suppose  for some $ 1 < i  < n$ that $\phi_{\Sigma_i}  \left(  x_i \cco{T_i^+}  \mathrm{g}    \mid   x_i \cc  A_i  \right)\geq c$ 
 \\
and $\phi_{\Sigma_i}  \left( \tilde x \cco{\tilde T^+}  \mathrm{g}  \mid  \tilde x \cc \tilde A  \right)\geq c$.  Then for all $h \leq h_0$ it holds that
$ \phi_{\Sigma_i}  \left(  \hat{\mathcal{R}}_i^*\mid  x_i \cc  \tilde x  \right) \geq c.
$
\end{proposition}

\noindent We first state the recent result that the random cluster model still has the RSW property at scales up to the correlation length. Here we need the wired boundary condition which is introduced in for example \cite{DC17}.

\begin{lemma}\label{dmtlem} (\cite{DCM20}, Lemma 8.5)
For any sufficiently large $C > 0$, there is an $\varepsilon > 0$ such that if $n \geq 1$ and $H \geq 0$ are such that $H n^2 \phi_{\Lambda_{n}, H=0}^{0, a=1} ( 0 \cc \partial \Lambda_{n}) \leq \varepsilon $ then 
\begin{align*}
\phi_{A_{n,2n}, H =0}^{1,a=1}( \Lambda_n  \not \cc \partial \Lambda_{2n}) \leq C \phi_{A_{n,2n}, H}^{1,a=1}( \Lambda_n  \not \cc \partial \Lambda_{2n}).
\end{align*} 
\end{lemma}
\noindent 
 
\noindent Translating the lemma into our setting yields the following lemma.  
\begin{lemma} \label{RSWh}
There exists a $h_0 > 0$ such that for $a \leq 1$  and $h \leq h_0$ it holds that
\begin{align*}
\phi^0_{\Sigma_i, \vec h} (  \hat{\mathcal{R}}_i^*) \geq c . 
\end{align*}
where $c > 0$ is (as always) independent of $0 < a \leq 1, \Sigma_i $ and  $ 0 \leq h < h_0$. 
Further, for any event $E$ depending only on edges in $D_i$ it holds that
\begin{align*}
\phi_{\Sigma_i}^0(  \hat{\mathcal{R}}_i^* \mid E) \geq c. 
\end{align*} 
\end{lemma}
\begin{proof}

If $n = \frac{2}{a}$ and $H = a^{\frac{15}{8}} h$ then using equation (1-arm)
\begin{align*}
    H n^2 \phi_{\Lambda_{n}, H=0}^{0, a=1} ( 0 \cc \partial \Lambda_{n}) \leq C (an)^\frac{15}{8} h = C 2^\frac{15}{8} h  \leq \varepsilon
\end{align*}
which can be satisfied by choosing $h$ sufficiently small (independent of $a$). Therefore

\begin{align*}
    \phi_{A_{2,4},h}^{1,a}(\Lambda_2  \not \cc \partial \Lambda_{4}) = 
    \phi_{A_{\frac{2}{a},\frac{4}{a}},h a^{\frac{15}{8}}  }^{1,1}(A_{ \frac{2}{a}}  \not \cc \partial A_{\frac{4}{a}} )
    \geq C \phi_{A_{\frac{2}{a},\frac{4}{a}},h=0 }^{1,1}(A_{ \frac{2}{a}}  \not \cc \partial A_{\frac{4}{a}} )
    = C \phi_{A_{2,4},h=0}^{1,a}(\Lambda_2  \not \cc \partial \Lambda_{4}).
\end{align*}

\noindent
Since $ \hat{\mathcal{R}}_i^*$ is decreasing it follows by (MON) and the RSW for usual rectangles \cite{DHN11} that 
\begin{align*}
\phi^{0,a}_{\Sigma_i,\vec h} (  \hat{\mathcal{R}}_i^*)  & \geq \phi^{0,a}_{\Sigma_i,h} (  \hat{\mathcal{R}}_i^*) \geq  \phi^{0,a}_{\Lambda,h}( \hat{\mathcal{R}}_i^*) \geq \phi^{0,a}_{\Lambda,h}(\Lambda_2(x_i)  \not \cc \partial \Lambda_{4}(x_i))  \geq \phi_{\Lambda,h}^{1,a}(\Lambda_2(x_i)  \not \cc \partial \Lambda_{4}(x_i)) \\
& \geq  \phi_{A_{2,4}, h}^{1,a}( \Lambda_2  \not \cc \partial \Lambda_{4}) \geq c \phi_{A_{2,4}, h =0}^{1,a}( \Lambda_2  \not \cc \partial \Lambda_{4}) \geq c.
\end{align*}
Using comparison between boundary conditions, and that because of the argument in Claim \ref{free_bc} removing all explored edges of the backbone acts as a free boundary condition it holds that 
\begin{align*}
\phi^0_{\Sigma_i,\vec h} (  \hat{\mathcal{R}}_i^* \mid E) \geq \phi^{1 \text{ on } \partial \Lambda_1(x_i), 0  \text{else}}_{\Sigma_i \backslash D_i, h} (  \hat{\mathcal{R}}_i^* ) \geq \phi_{A_{2,4}, h}^{1,a}( \Lambda_2  \not \cc \partial \Lambda_{4}) \geq c.
\end{align*}
\end{proof}

\noindent Next, we need the general result that we can do mixing also with a magnetic field at scales up to the correlation length and that we,  up to constants, can decorrelate events in $T_i $ from events in $(T_i \cup \Lambda_4(x_i))^c$.  Define $J_i = (T_i \cup \Lambda_4(x_i) ) \backslash \bar \gamma_i $ and $ \tilde J = ( \tilde T \cup \Lambda_4(\tilde x) ) \backslash \bar  \gamma_x $. 
Define also  $\hat{\mathcal{R}}^*_\sim$  similarly to  $ \hat{\mathcal{R}}^*_i$ as an event in the vicinity of $\tilde x$ instead of $x_i$. 
\begin{lemma} \label{mix} 
(Mixing) Let $E_1, E_2$ be increasing events that only depend on edges in the boxes $T_i \cup \Lambda_2(x_i)$ and $ \tilde T \cup \Lambda_2(\tilde x)$ respectively. Then for $1 \leq i \leq n-2$ it holds that 
\begin{align*}
\phi^0_{\Sigma_i}( E_1 \cap E_2) \asymp \phi^{0,a}_{J_i,  \vec h}(E_1)\phi^{0,a}_{\tilde J,  \vec h}(E_2)  
\end{align*}
 Similarly, if $l\leq 1$, $z_1,z_2$ are such that $ \Lambda_{2l}(z_1) \cap  \Lambda_{2l}(z_2)  = \emptyset$ and $E_1, E_2$ are increasing events that only depend on edges in the boxes $\Lambda_l(z_1), \Lambda_l(z_2)$ respectively. Then
\begin{align*}
\phi^0_{\Sigma_i}( E_1 \cap E_2) \asymp \phi^{0,a}_{\Lambda_{2l} (z_1) \backslash ( \bar \gamma_i \cup \bar \gamma_x),  \vec h}(E_1)\phi^{0,a}_{\Lambda_{2l} (z_2) \backslash ( \bar \gamma_i \cup \bar \gamma_x),  \vec h}(E_2). 
\end{align*}
\end{lemma}
\begin{proof}
We prove the first statement first. Define  $E = E_1 \cap E_2$. 
It follows from Lemma \ref{RSWh} that 
\begin{align*}
c \phi^0_{\Sigma_i} (E) \leq  \phi_{\Sigma_i} (E \mid  \hat{\mathcal{R}}_i^* )
\end{align*}
and similarly we can condition on $\Rt$. 
Using that closed dual paths inside the annulus give rise to monotonicity properties as free boundary conditions (which is for example proven in Lemma 11 in \cite{cjn20}) we obtain that
\begin{align*}
 c  \phi^0_{\Sigma_i}( E) & \leq \phi_{\Sigma_i}( E_1, E_2 \mid  \hat{\mathcal{R}}_i^*, \Rt ) \leq \phi^0_{J_i \cup \tilde J, \vec h}(E_1, E_2 ) = \phi^0_{J_i , \vec h}(E_1)\phi^0_{\tilde J, \vec h}(E_2).
\end{align*}
Since the reverse inequality is (FKG) and (MON) the first result follows. 
The second assertion follows mutatis mutandis using that the estimates from the proof of Lemma \ref{RSWh} which by Lemma \ref{dmtlem} also work on smaller scales and using the event $\{\partial \Lambda_l(z_i) \not \cc \partial \Lambda_{2l}(z_i) \}$ for $i =1,2$ instead of $ \hat{\mathcal{R}}_i^*$ and  $ \Rt$.
\end{proof}

\noindent With the lemmas proven we continue to the proof of the Proposition  \ref{prop2}. 

\begin{proof} [Proof of Proposition \ref{prop2}]
{First, notice that 
\begin{align*}
\phi_{\Sigma_i}  \left( \hat{\mathcal{R}}_i^* \mid x_i \cco{\Sigma_i^+}  \tilde x \right) \geq  
\phi_{\Sigma_i} \left( \hat{\mathcal{R}}_i^* \mid   x_i \cco{T_i} \mathrm{g}, \tilde x \cco{\tilde T}  \mathrm{g} \right) \phi_{\Sigma_i}  \left( x_i \cco{T_i}  \mathrm{g}, \tilde x \cco{\tilde T} \mathrm{g}   \mid x_i \cco{\Sigma_i^+} \tilde x \right). 
\end{align*}
From the mixing argument in Lemma \ref{RSWh} it follows that  $\phi_{\Sigma_i} \left( \hat{\mathcal{R}}_i^*\mid   x_i \cco{T_i} \mathrm{g}, \tilde x \cco{\tilde T} \mathrm{g} \right) \geq c$. 
Thus, we just need to prove that 
$
\phi_{\Sigma_i} \left(  x_i \cco{T_i}  \mathrm{g},  \tilde x \cco{\tilde T}  \mathrm{g}  \right) \geq c \phi_{\Sigma_i} \left( x_i  \cco{\Sigma_i^+} \tilde x \right). 
$
Notice that 
\begin{align*}
  \{  x_i  \cco{\Sigma_i^+} \tilde x \}  \subset  \left(   \{  x_i \cco{T_i}  \mathrm{g} \}  \cup   \{  x_i \cc A_i  \}  \right)  \cap 
 \left( \{  \tilde x \cco{\tilde T}  \mathrm{g} \} \cup   \{  \tilde x \cc \tilde A \}  \right).
\end{align*}
Now, by first a union bound, then (Mixing) and the assumption and finally (MON) and (FKG)
\begin{align*}
\phi_{\Sigma_i}  \left( x_i  \cco{\Sigma_i^+} \tilde x \right) &\leq \phi_{\Sigma_i} \left(  x_i \cco{T_i}  \mathrm{g} , \tilde x \cco{\tilde T}  \mathrm{g}  \right) + \phi_{\Sigma_i} \left(   x_i \cco{T_i}  \mathrm{g},  \tilde x \cc  \tilde A  \right)
\\ &+\phi_{\Sigma_i} \left(   x_i \cc A_i, \tilde x \cco{\tilde T} \mathrm{g} \right) + \phi_{\Sigma_i} \left(    x_i \cc A_i,  \tilde x \cc \tilde A \right) \\
& \leq 4 C \phi^{0,a}_{J_i, \vec h} \left( x_i \cco{T_i}  \mathrm{g} \right)  \phi^{0,a}_{\tilde J, \vec h}  \left( \tilde x \cco{\tilde T}  \mathrm{g} \right)  
\leq c \phi_{\Sigma_i} \left(  x_i \cco{T_i}  \mathrm{g} ,  \tilde x \cco{\tilde T}  \mathrm{g} \right).
\end{align*} 
}
\end{proof}

\noindent We now turn to the main technical part for proving Proposition \ref{prop5}. 
 From now on, we will  assume, for notational reasons,  without loss of generality that $x_i$ is on the right side of the inner boundary of the annulus $A_{9i, 9(i+1)}$, i.e. that $x_i = (\frac{9i}{2}, y) $ for some $y$ such that $ - \frac{9i}{2} \leq y \leq \frac{9i}{2}$. Then define $L = \lbrack 1, 2 \rbrack \times \lbrack 0,  \frac{1}{10} \rbrack + x_i$ and $\tilde L$ similarly to be a rectangle in the vicinity of $\tilde x$.

 \begin{lemma} \label{prison}
For each $ 1 < i <  n$ it holds that
\begin{align*}
\phi^0_{\Sigma_i} ( x_i \cc L \mid x_i \cc A_i) \geq c
\text{  as well as } 
\phi^0_{\Sigma_i} ( \tilde x  \cc L \mid  \tilde x \cc \tilde A) \geq c. 
\end{align*}
\end{lemma}

\noindent Let $C(x_i)$ be number of points in $\Lambda_2(x_i)$ connected to $x_i$ without using edges to the ghost and $N_{M,i}  = \{ C(x_i) \geq M \}$ for each $M \in \N$.

\begin{lemma} \label{bigclust}
Let $k > 0$. Then, 
\begin{align*}
 \phi^0_{\Sigma_i} \left( N_{  k  a^{-\frac{15}{8}},i } \mid x_i \cc L,  x_i \cc  A_i \right)  \geq c.
 \end{align*}
\end{lemma}

\noindent  Then let us start proving the lemmas.  To do that we need to show that crossings of topological rectangles exist with constant probability. This is done in (\cite{CDH13}, Theorem 1.1) if  the discrete extremal length $l_{\Omega}\lbrack (ab), (cd) \rbrack $ is bounded.  From  (\cite{CDH13},(3.7)) (see also \cite{Che12}) we have the following characterisation of the discrete extremal length 
\begin{align*}
l_{\Omega}( (ab), (cd)) =  \sup_{g: E(\Omega) \to \R_+  \cup \{0\} } \frac{ \left(\inf_{ \gamma: (ab) \cc (cd)} \sum_{e \in \gamma} g_e \right)^2 }{\sum_e g_e^2}
\end{align*}
where the supremum is over all non-negative, not identically zero functions on the edges. Using this representation we obtain the following lemma (which is equivalent to Rayleigh's monotonicity law) . 
\begin{lemma} \label{chelkak}
Let $T$ be a topological rectangles with marked points $(abcd)$. Let $e,f$ be points on $(ad)$ and $\gamma$ a path from $e$ to $f$ inside $T$. Let $\tilde T$ be the points in $T$ reachable from $b$ in $T \backslash \gamma$. Note $\tilde T$ is a new topological rectangle with marked points $(abcd)$.
Then, 
\begin{align*}
l_{T} \left( (ad), (bc) \right) \geq l_{\tilde T} \left( (ad), (bc) \right) &&  \text{   as well as }  &&  l_{T} \left( (ab), (cd) \right) \geq l_{\tilde T} \left( (ab), (cd) \right). 
\end{align*}
\end{lemma}
\begin{proof}
We prove the first inequality, the second follows similarly. Since the graph $\tilde T$ is finite the supremum and infimum are attained and we get some maximizing function $\tilde g$ for $\tilde T$. Now, define the function $g$ by extending $ \tilde g$  with $g(e) = 0 $ whenever $ e \not \in T \backslash \tilde  T$. Then, 
\begin{align*}
l_{ \tilde T}( (ad) \cc (bc) )  = \frac{ \left(\inf_{ \gamma: (ad) \cc (bc) \text{ in } \tilde T} \sum_{e \in \gamma} \tilde g_e \right)^2 }{\sum_e  \tilde g_e^2} = \frac{ \left(\inf_{ \gamma: (ad) \cc (bc) \text{ in } T} \sum_{e \in \gamma} g_e \right)^2 }{\sum_e  g_e^2} \leq l_{T}( (ad) \cc (bc) ). 
\end{align*}
The second equality follows since any path $\gamma: (ad) \cc (bc) \text{ in } T$  has a subpath $ \tilde \gamma: (ad) \cc (bc) \text{ in } \tilde  T$. 
The inequality follows since the function $g$ is just one element in the supremum defining the discrete extremal length. 
\end{proof} 

\noindent Using Lemma \ref{chelkak} we can now prove Lemma \ref{prison}

\begin{figure}

  \centering
  \includegraphics[width=10cm]{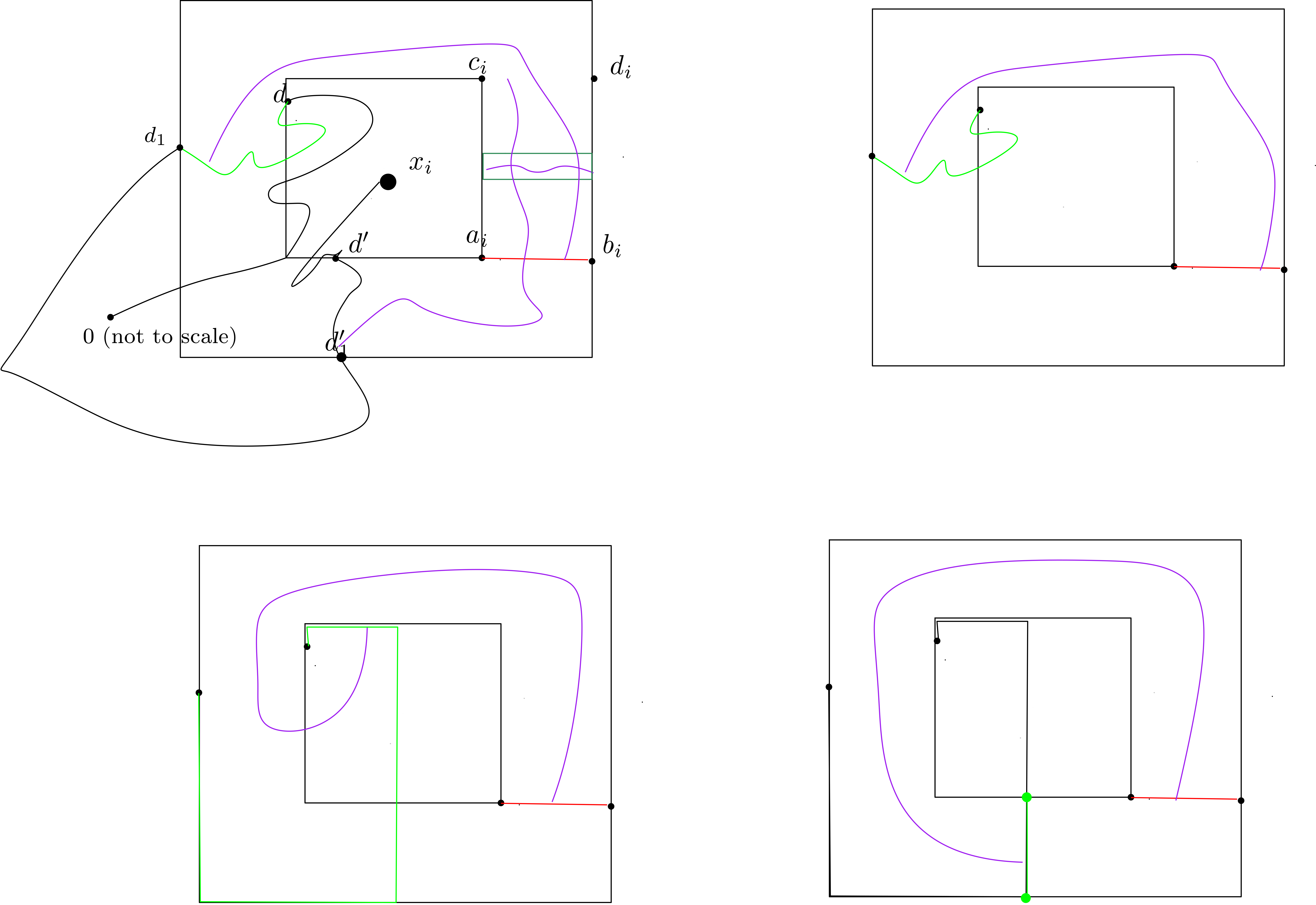}
    \caption{a) The backbone and the paths constructed in the steps in the proof of Lemma \ref{prison}.
    b) The path $(a_i b_i) \cc (d d_1)$ is (strictly) inside the domain $T_{i,1}$. c) The domain $T_{i,2}$ and the path  $(a_i b_i)  \cc (d d_1)$. 
    d) The domain $T_{i,3}$ and an example of the path $(a_i b_i) \cc (c_i d_i)$ \label{brew}}
\end{figure}

 \begin{proof}[Proof of Lemma \ref{prison}]
Define the explored vertices $\mathcal{V}$ of the backbone to be all vertices with at least one incident explored edge. 
Then define $U$ to be the set of vertices in $V(\Lambda) \backslash ( \mathcal{V} \cup D_i) $ with at least one edge to  $\mathcal{V}$. Since $ \gamma_i \cap \Lambda_{2,R}(x_i) = \emptyset$ there exists at least one $*$-path (i.e. a path that can also jump diagonally) ${P_i}^*$  in $U$ from $d$ to a vertex in $\partial \Lambda_2^c(x_i) $. 
From such a $*$-path  $P_i^*$ we can construct a usual path $P_i$ just going around the plaquette every time ${P_i}^*$ jumps diagonally. Let the first vertex that $P_i$ hits in $\Lambda_2(x_i)^c$ be $d_1$ and denote henceforth the path $P_i$ by $(d d_1)$. Define $d_1' $ similarly following the outside of the backbone from $d'$.

Now, let $\Lambda_1(x_i)_R$ denote the right half of the box $\Lambda_1(x_i)$. Define $a_i = x_i + (1,-1)$, $b_i = x_i + (2,-1)$, $a_i' = x_i' + (1,1)$ and $b_i' = x_i + (2,1)$ see Figure \ref{brew}. Define $T_{i,1} =\Lambda_{2}(x_i) \backslash D_i $.  Then let $\mathcal{S}_i \in \{0,1 \}^{ E( T_{i,1} )} $ be the event defined by
\begin{align*}
\mathcal{S}_i  =  \left \{ (a_i b_i)   \cco{T_{i,1} } (d d_1) \right \} \bigcap \left \{ (a_i' b_i')   \cco{T_{i,1} } (d' d_1') \right \} \bigcap \left \{ (a_i a_i') \cco{L} (b_i b_i')  \right \}. 
\end{align*}
I.e. $\mathcal{S}_i$ ensures that any path from $x_i$ to $\Lambda_{9i+1}^c$ will intersect a cluster of open edges that in particular hits $L$. We claim that 
\begin{claim}
$\phi_{\Sigma_i}^0( \mathcal{S}_i) \geq c$.
\end{claim} 
\begin{proof} 
 We prove that each of the three events defining $\mathcal{S}_i $ has a positive probability. 
 That  
 \begin{align*}
 \phi_{\Sigma_i}^0 \left( (a_i a_i') \cco{L} (b_i b_i') \right) \geq c
 \end{align*} follows from RSW for usual rectangles \cite{DHN11}. Thus, by symmetry it suffices to prove 
 \begin{align*}
\phi_{\Sigma_i}^0(   (a_i b_i)   \cco{T_{i,1}} (d d_1) ) \geq c.
 \end{align*}
Notice that the path $(d d_1)$ does not leave the left half of the box $\Lambda_2(x_i)$ since the backbone is only in the left half. If we consider a new topological rectangle $T_{i,2}$ to be $\Lambda_2(x_i)_L$ union the top-right quarter of $A_{1,2}(x_i)$ with the four marked points $a_i, b_i , d, d_i$ and where we use the part of  $\partial \Lambda_1(x_i)$ from $x_i+(0,1)$ until $d$ as the boundary twice as shown on Figure \ref{brew}. Then the path $(d d_1)$ has the form of $\gamma $ in  Lemma \ref{chelkak}  so we conclude that 
\begin{align*}
  l_{T_{i,1}} \left(  (a_i b_i), (d d_1) \right) 
\leq    l_{T_{i,2}} \left(  (a_i b_i), (d d_1) \right) 
\end{align*}
Define $c_i = x_i + (0,-1)$ and $d_i = x_i + (0,-2)$. Then $c_i, d_i$ are on the segment $(d d_1)$ and thus
\begin{align*}
  l_{T_{i,2}} \left(  (a_i b_i), (d d_1) \right) \leq l_{T_{i,2}} \left(   (a_i b_i), (c_i d_i) \right) \leq l_{T_{i,3}}  \left(  (a_i b_i) , (c_i d_i) \right) \leq c 
  \end{align*}
where we in the last step considered a new topological rectangle $T_{i,3}$ where we used the part of $A_{1,2}(x_i)$ enclosed by $(a_i b_i) $ and $ (c_i d_i) $ as shown on Figure \ref{brew} which has bounded discrete extremal length. Therefore $l_{T_{i,1}} \left(  (a_i b_i), (d d_1) \right) \leq c$ which means by  (\cite{CDH13}, Theorem 1.1) if that 
\begin{align*}
\phi_{\Sigma_i}^0 \left(  (a_i b_i)   \cco{T_{i,1} } (d d_1)  \right)   \geq \phi_{T_{i,1}, \vec h}^{0,a} \left(  (a_i b_i)   \cco{T_{i,1} } (d d_1) \right)  \geq c. 
\end{align*} That $\phi_{\Sigma_i}^0( \mathcal{S}_i) \geq c$ then follows from (FKG). 
\end{proof}
\noindent Now, to finish the proof of the lemma note that since 
$ \{ x_i \cc A_i \} \cap \mathcal{S}_i \subset  \{x_i \cc L\} $ then by (FKG) 
\begin{align*}
\phi_{\Sigma_i}^0 \left( x_i \cc L \mid x_i \cc A_i \right)  \geq \phi_{\Sigma_i}^0 \left(  \mathcal{S}_i \mid x_i \cc A_i \right)  \geq \phi_{\Sigma_i}^0\left(  \mathcal{S}_i \right)  \geq c. 
\end{align*} 
\end{proof}

\noindent We end by proving Lemma \ref{bigclust}. 

\begin{proof} [Proof of Lemma \ref{bigclust}]
We use some ideas from Lemma 3.1 in  \cite{CGN15}. 
Consider a square $B = \Lambda_{\frac{1}{10}}(\frac{3}{2}, \frac{1}{20}) + x_i$ corresponding with the previous lemma such that $L$ passes through $B$.
Define the event $\mathcal{S}$ to be $\mathcal{S}_i$ where there is also a crossing of each of the four (overlapping) rectangles that make up the annulus $A_{\frac{1}{10}, \frac{2}{10}}(\frac{3}{2}, \frac{1}{20})$ around $B$ as shown on Figure \ref{ghost}. 
By RSW for usual rectangles and (FKG) we know that $ \phi^0_{\Sigma_i}( \mathcal{S}) \geq c$. 
By the definition of $\mathcal{S}_i$ from Lemma \ref{prison} $\{x_i  \cc L \} \cap \mathcal{S} = \{ x_i \cc B \} \cap \mathcal{S}$ and so by (FKG) and Lemma \ref{prison} we get that 
\begin{align*}
\phi^0_{\Sigma_i} \left( x_i  \cc B \mid  x_i  \cco{T_i} A_i  , \mathcal{S} \right) \geq \phi^0_{\Sigma_i} \left( x_i  \cc L \mid  x_i  \cco{T_i} A_i  , \mathcal{S} \right) \geq c. 
\end{align*}
\noindent Now, since
\begin{align*}
 \phi^0_{\Sigma_i} \left( N_{  k  a^{-\frac{15}{8}}, i } \mid x_i  \cc L,  x_i   \cco{T_i} A_i   \right)   \geq  \phi^0_{\Sigma_i} \left( N_{  k  a^{-\frac{15}{8}}, i } \mid x_i  \cc B, x_i  \cco{T_i} A_i  , \mathcal{S} \right) \phi^0_{\Sigma_i} \left( x_i  \cc B \mid  x_i  \cco{T_i} A_i  , \mathcal{S} \right) 
\end{align*}
it suffices to prove that 
\begin{align*}
 \phi^0_{\Sigma_i} \left( N_{  k  a^{-\frac{15}{8}}, i } \mid x_i  \cc B, x_i   \cco{T_i} A_i  , \mathcal{S} \right) \geq c. 
\end{align*}
This follows from a second moment estimate. 
So let $N_B = \sum_{z \in B} 1_{z \cc x_i} $. 
First using the fact that $ \mathcal{S} \cap \{x_i \cc L \} \cap \{ z \cc \partial \Lambda_{\frac{3}{10}}(x) \} \subset \{x_i \cc z \}$ and then by (FKG), (MON) and (1-arm) we get 
\begin{align*}
\mathbb{E}(N_B \mid  x_i \cc B, x_i  \cco{T_i} A_i, \mathcal{S}) 
& =  \sum_{z \in B} \phi^0_{\Sigma_i} \left( x_i  \cc z \mid   x_i  \cc B,   x_i  \cco{T_i} A_i  , \mathcal{S} \right) \\
&  \geq  \sum_{z \in B} \phi^0_{\Sigma_i} \left( z \cc  \partial \Lambda_{\frac{3}{10}}(z) \mid   x_i  \cc B,   x_i  \cco{T_i} A_i  , \mathcal{S} \right) \\
&   \geq  \sum_{z \in B} \phi^0_{\Lambda_{\frac{3}{10}}(z)} \left( z \cc  \partial \Lambda_{\frac{3}{10}}(z) \right)
= \frac{1}{100 a^2} C a^{\frac{1}{8}} = c a^{-\frac{15}{8}}. 
\end{align*}

\begin{figure}

  \centering
  \includegraphics[width=8cm]{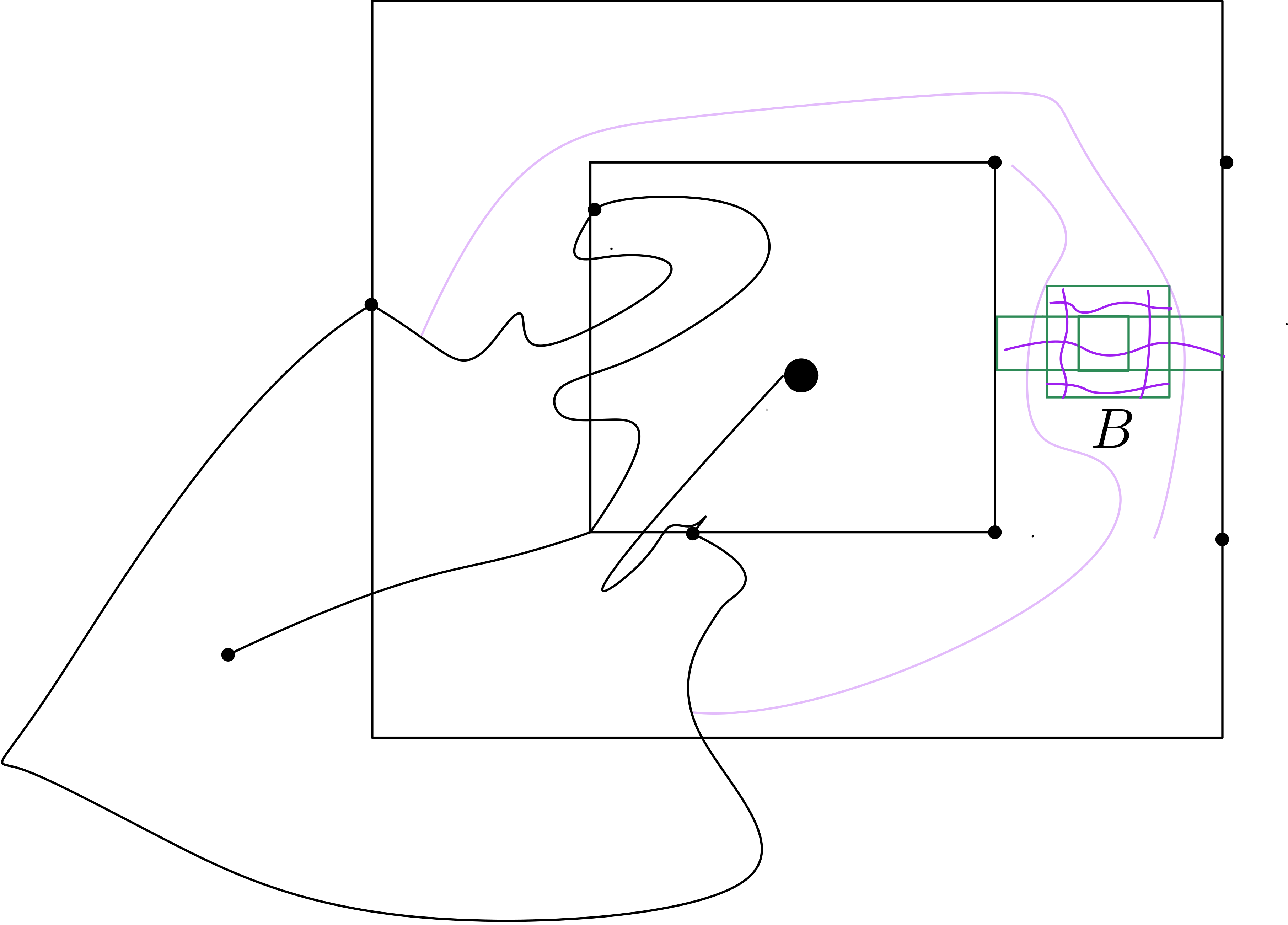}
    \caption{To the right we sketch the situation in Lemma \ref{bigclust}. We keep the paths that we have seen exist in Lemma \ref{prison} with constant probability. Then we use RSW to construct some more paths encircling the box $B$ as shown. Later in the proof we split the box $B$ up into dyadic annuli. \label{ghost}}
\end{figure}

\noindent   Let us then consider the second moment. First we use that  $ \phi^0_{\Sigma_i}(x_i  \cc B \mid x_i  \cco{T_i} A_i  ) \geq c$,  $ \phi^0_{\Sigma_i}(\mathcal{S}) \geq c $ and then we do a dyadic summation for each $z$ partitioning $B$ into annuli
$ A_{2^{k-1}, 2^k}(z)$  for $k$ such that $ - m \leq k \leq 0$ where $m \in \N$ is chosen such that $A_{2^{k-1}, 2^k}(x) = \emptyset $ for $k < - m$. 
\begin{align*}
& \mathbb{E}(N_B^2 \mid  x_i \cc B,  x_i  \cco{T_i} A_i , \mathcal{S}) 
=  \sum_{z,y \in B} \phi^0_{\Sigma_i} \left( x_i  \cc z, x_i  \cc y \mid   x_i  \cc B,  x_i  \cco{T_i} A_i  , \mathcal{S} \right) \\
& \leq \frac{c}{ \phi^0_{\Sigma_i} (  x_i  \cco{T_i} A_i )} \sum_{z,y \in B}  \phi^0_{\Sigma_i} \left( x_i  \cc z, x_i  \cc y, x_i  \cco{T_i} A_i   \right) \\
&\leq  \frac{c}{ \phi^0_{\Sigma_i} (  x_i  \cco{T_i} A_i )} \sum_{z  \in B} \sum_{k = -m}^0 \sum_{y \in A_{2^{k-1}, 2^k}(z)}   \phi^0_{\Sigma_i} \left(z \cc \partial \Lambda_{2^{k-2}}(z) , y \cc \partial \Lambda_{2^{k-2}}(y) , x_i  \cc \partial \Lambda_{\frac{1}{20},\frac{1}{20}}(x_i) \right) \\
&\leq  \frac{c  \phi^0_{\Sigma_i}(x_i  \cc \partial \Lambda_{\frac{1}{20},\frac{1}{20}}(x_i) ) }{ \phi^0_{\Sigma_i} (  x_i  \cco{T_i} A_i )} \sum_{z  \in B} \sum_{k = -m}^0 \sum_{y \in A_{2^{k-1}, 2^k}(z)}  \phi^0_{\Sigma_i} \left( z \cc \partial \Lambda_{2^{k-2}}(z) \right)\phi^0_{\Sigma_i} \left(  y \cc \partial \Lambda_{2^{k-2}}(y) \right) \\
&\leq  \sum_{z  \in B} \sum_{k = -m}^0 c_1 \left( \frac{2^k}{a} \right)^2  \left( \frac{a}{2^{k-2}} \right)^{\frac{2}{8}} = c_2 a^{-2}  a^{-2} a^\frac{1}{4} \sum_{k = -m}^0 2^{2k - \frac{k}{4}} = c_3 a^{- \frac{15}{4}}
\end{align*}
where we also used (1-arm) several times and that (Mixing) holds at all scales smaller than our fixed macroscopic scale. The conclusion follows from the Paley-Zygmund inequality 
\begin{align*}
\phi^0_{\Sigma_i} \left( N_{  k  a^{-\frac{15}{8}}, i } \mid x_i  \cc B, x_i   \cco{T_i} A_i  , \mathcal{S}  \right)  \geq c a^{\frac{15}{4} - \frac{15}{4}} = c. 
\end{align*}
\end{proof}

Finally, let us prove Proposition \ref{prop5}. 

\begin{proof}  [Proof of Proposition \ref{prop5}] 
By Proposition \ref{prop2} it suffices to prove  $\phi^0_{\Sigma_i} \left(  x_i \cco{T_i}  \mathrm{g}  \mid  x_i \cc A_i \right)\geq c$ and the similar inequality for $\tilde T$ and $\tilde x$ which will follow in the same way. 
First, notice that by Lemma \ref{prison} and Lemma \ref{bigclust}
\begin{align*}
\phi^0_{\Sigma_i} \left( N_{  k  a^{-\frac{15}{8}}, i } \mid   x_i \cc A_i  \right) \geq 
\phi^0_{\Sigma_i} \left( N_{  k  a^{-\frac{15}{8}}, i } \mid x_i \cc L,   x_i \cc A_i \right) 
\phi^0_{\Sigma_i} \left(x_i \cc L \mid  x_i \cc A_i  \right) \geq c \cdot c   
\end{align*}
Now, we can make the following observation using Lemma \ref{ghosted} to find
\begin{align*}
\phi^0_{\Sigma_i}(  x_i \cco{T_i}  \mathrm{g}  \mid   x_i \cc A_i) 
&\geq \phi^0_{\Sigma_i} \left(N_{  k  a^{-\frac{15}{8}}, i },  x_i \cco{T_i}  \mathrm{g}  \mid x_i \cc A_i \right) \\
& =  \phi^0_{\Sigma_i} \left(  x_i \cco{T_i}  \mathrm{g}   \mid N_{  k  a^{-\frac{15}{8}}, i },  x_i \cc A_i \right) \phi^0_{\Sigma_i} \left( N_{  k  a^{-\frac{15}{8}}, i } \mid  x_i \cc A_i \right) \\
& \geq  \tanh \left( k  a^{-\frac{15}{8}} h a^{\frac{15}{8}} \right) c \geq c(k,h). 
\qedhere
\end{align*}
\end{proof}

\section*{Acknowledgments}
The authors would like to thank Wendelin Werner for establishing collaboration and support through the SNF Grant 175505. 
The first author would like to thank the Swiss European Mobility Exchange program as well as the Villum Foundation for support through the QMATH center of Excellence(Grant No. 10059) and the Villum Young Investigator (Grant No. 25452) programs.  Further, thanks to Hugo Duminil-Copin and Ioan Manolescu for discussions.

\bibliographystyle{unsrt}

\bibliography{Expdecay.bib}

\end{document}